\documentclass{article}
\usepackage[a4paper]{geometry}
\usepackage[utf8]{inputenc}
\usepackage{url}
\usepackage{graphicx}
\usepackage[space]{grffile}
\usepackage{amsmath}
\usepackage{amsfonts}
\usepackage[colorlinks=true, citecolor=blue, linkcolor=blue, urlcolor=blue]{hyperref}
\usepackage{xcolor}
\usepackage{textgreek}
\usepackage{todonotes}
\usepackage{subcaption}
\usepackage{multirow}
\usepackage{marvosym}
\usepackage{booktabs}

\usepackage{caption}
\DeclareCaptionLabelFormat{adja-page}{\hrulefill\\#1 #2 \emph{(previous page)}}

\newcommand{\red}[1]{#1}
\newcommand{\unit}[1]{\,\text{#1}} 
\newcommand\orcid[1]{\hspace{0.03cm}\href{https://orcid.org/#1}{\includegraphics[scale=0.05]{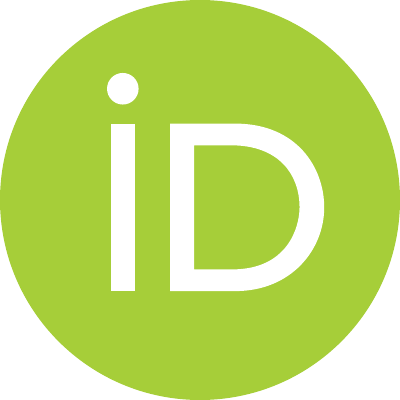}}\hspace{0.05cm}}

\usepackage[numbers]{natbib}
\title{Plastic Arbor: a modern simulation framework for\\synaptic plasticity -- from single synapses to networks of morphological neurons}
\author{Jannik Luboeinski\orcid{0000-0002-1959-5644}\textsuperscript{1,2,3,\Letter}, Sebastian Schmitt\orcid{0000-0002-7935-0470}\textsuperscript{1,2}, Shirin Shafiee\orcid{0009-0009-7271-8939}\textsuperscript{1,2},\\Thorsten Hater\orcid{0000-0002-6249-7169}\textsuperscript{4}, Fabian Bösch\orcid{0009-0002-3236-8187}\textsuperscript{5}, Christian Tetzlaff\orcid{0000-0003-1901-6232}\textsuperscript{1,2,3}}

\date{
    {\small
    \textsuperscript{1} III. Institute of Physics -- Biophysics, University of Göttingen, Göttingen, Germany\\
    \textsuperscript{2} Department for Neuro- and Sensory Physiology, University Medical Center Göttingen,\\ Göttingen, Germany\\
    \textsuperscript{3} Campus Institute Data Science (CIDAS), Göttingen, Germany\\
    \textsuperscript{4} Jülich Supercomputing Centre, Forschungszentrum Jülich, Jülich, Germany\\
    \textsuperscript{5} Swiss National Supercomputing Centre, ETH Zürich, Zürich, Switzerland\\
    \smallskip
    \textsuperscript{\Letter} Correspondence: jannik.luboeinski@med.uni-goettingen.de\\}
    \vskip 1cm
    29 August 2025}

\begin{document}

\maketitle
\vspace{-0.7cm}
\addcontentsline{toc}{section}{Abstract}
\section*{Abstract}
Arbor is a software library designed for efficient simulation of large-scale networks of biological neurons with detailed morphological structures. It combines customizable neuronal and synaptic mechanisms with high-performance computing, supporting multi-core CPU and GPU systems.

In humans and other animals, synaptic plasticity processes play a vital role in cognitive functions, including learning and memory. Recent studies have shown that intracellular molecular processes in dendrites significantly influence single-neuron dynamics. However, for understanding how the complex interplay between dendrites and synaptic processes influences network dynamics, computational modeling is required.

To enable the modeling of large-scale networks of morphologically detailed neurons with diverse plasticity processes, we have extended the Arbor library to support simulations of a large variety of spike-driven plasticity paradigms. To showcase the features of the extended framework, we present examples of computational models, beginning with single-synapse dynamics, progressing to multi-synapse rules, and finally scaling up to large recurrent networks. While cross-validating our implementations by comparison with other simulators, we show that Arbor allows simulating plastic networks of multi-compartment neurons at nearly no additional cost in runtime compared to point-neuron simulations. In addition, we demonstrate that Arbor is highly efficient in terms of runtime and memory use as compared to other simulators.

Using the extended framework, as an example, we investigate the impact of dendritic structures on network dynamics across a timescale of several hours, finding a relation between the length of dendritic trees and the ability of the network to efficiently store information. 

By our extension of Arbor, we aim to provide a valuable tool that will support future studies on the impact of synaptic plasticity, especially, in conjunction with neuronal morphology, in large networks.

\addcontentsline{toc}{section}{Author Summary}
\section*{Author Summary}

Understanding how synaptic plasticity shapes the dynamics of the brain requires simulation tools that provide efficient, scalable computation to integrate detailed neuronal morphology into large networks of neurons. Therefore, we extended the open‑source Arbor simulator to enable modeling a comprehensive range of spike‑driven plasticity mechanisms. Using progressively more complex examples – from a single synapse to recurrent networks of thousands of morphological neurons – we demonstrate that the extended Arbor framework matches the results of established simulators. At the same time, it uncloses a regime of new investigations. Benchmarking on modern hardware shows that adding morphology incurs only modest runtime and memory overhead as compared to point-neuron models. Moreover, a comparison with the widely used NEURON framework shows that Arbor is markedly more efficient, even when computing additional plasticity dynamics. Leveraging this performance, as an example, we explore how dendritic length and cell size influence long‑term memory recall, revealing that larger dendritic trees can impair, whereas larger cell diameters can enhance, pattern‑completion performance. The extended Arbor framework and the considered examples of plasticity models are freely available and shall serve to provide a powerful platform for researchers investigating the interplay of cellular structure, synaptic plasticity, and network dynamics.


\section{Introduction}

Over the past decades, a number of open-source software simulators have been developed to facilitate the investigation of biological neural networks. Current prominent examples are (Core) NEURON \cite{Hines1989,Kumbhar2019}, NEST \citep{GewaltigDiesmann2007}, and Brian 2 \citep{Stimberg2019}. 
While NEST and Brian 2 are widely used for the simulation of large-scale networks of point neurons, NEURON is a well-established tool for modeling neurons with detailed morphology, with its first version already released in the 1980s \cite{Hines1989}. Notably, NEURON enables to flexibly create realistic neuron models in its own script language called NMODL. 
Over the past decades, NEURON was extended to include more features as well as to keep up with the fast development in computing hardware \cite{Kumbhar2019,Awile2022}. This has, however, yielded a complex code base that constrains usability, flexibility, and the optimization for modern hardware backends.
To overcome these limitations, Arbor has been developed, which is a new simulator library for networks of neurons with detailed morphology. With a Python frontend and support for various model-descriptive formats, including NMODL, Arbor facilitates the implementation and customization of neuron and synapse models.
At the same time, Arbor offers heavily optimized execution on different hardware systems \cite{AbiAkar2019}. It supports, in particular, modern backend architectures such as multi-core central processing units (CPUs), graphics processing units (GPUs), and message passing interface (MPI) ranks \citep{AbiAkar2019}, and is openly developed and available on GitHub \cite{ArborCoreCode}. Besides CoreNEURON, Arbor is the only simulator with comprehensive GPU and MPI support for multi-compartment models.

Synaptic plasticity is the ability of synapses to strengthen or weaken in response to neural activity, which is essential for learning and memory \cite{Martin2000,Abraham2019}. So-called long-term synaptic plasticity is crucial for both short- and long-term memory and appears in two types: long-term potentiation (LTP) and long-term depression (LTD), which strengthen and weaken synaptic connections, respectively. These processes mainly involve changes in the distribution of postsynaptic receptors and in the structure of dendritic spines, driven by complex biochemical and biophysical mechanisms within synapses as well as in dendrites \cite{Smolen2012,Gallimore2018,MakiMarttunen2020,BeckerTetzlaff2021,BonillaQuintana2021}. Their dysregulation has further been linked to neurological disorders like Alzheimer's disease and schizophrenia \cite{Berridge2013,BonillaQuintana2022}. On the other hand, recent machine learning approaches also make use of synaptic and dendritic processes to improve the performance of large-scale neural networks \cite{acharya2022dendritic,pagkalos2024leveraging,zheng2024temporal}. Thus, investigations on the functional impact of synapses and dendrites on network dynamics are becoming more and more fundamental to advance research in areas of neuroscience, medicine, and machine learning.   

Arbor offers a powerful tool for capturing processes at the synaptic, dendritic, and neuronal levels, enabling the examination of their links to network dynamics.
Although the core functionality for running network simulations has been available since the inception of the Arbor project \citep{AbiAkar2019}, the mechanisms required for modeling plasticity-related processes have largely been absent.
In our present work, we fill this gap extending the Arbor simulator by the general functionality needed to model diverse spike-driven plasticity rules (e.g., \cite{GraupnerBrunel2012,HirataniFukai2017,LuboeinskiTetzlaff2021}), which constitutes the `Plastic Arbor' framework. 
By this, we provide a basis for investigating the functional impact of plasticity dynamics in \emph{large} networks of morphological neurons via our publicly available code (see the data availability statement below, or visit \href{https://github.com/tetzlab/plastic_arbor}{https://github.com/tetzlab/plastic\_arbor}).

In this article, we aim to present Arbor's new set of features to model plasticity in large-scale networks of multi-compartment neurons, as well as to demonstrate that Arbor enables highly efficient simulation of such networks in terms of runtime and memory use. 
We describe the novel functionality in two different sections, where we present methods and results in a side-by-side fashion.
The new technical features needed to implement plasticity rules are described in the section `Design and Implementation: Extensions of the Arbor core code'. 
In the section `Results: Computational modeling with synaptic plasticity', we present examples for the newly implemented technical features introduced before, with increasing complexity of the implemented models.
We start from the level of simulating a widely-used spike-timing-dependent plasticity rule in a single synapse, then consider plasticity rules involving multiple synapses, move on to more complex plasticity rules including several hidden states, and finally reproduce findings from large recurrently connected networks.
We cross-validate all Arbor implementations with the Brian 2 simulator \citep{Stimberg2019} or with model-specific custom simulators. 
Importantly, the newly extended functionality of Arbor has enabled us to provide first predictions about the network dynamics underlying synaptic memory consolidation across networks of neurons of different morphology. We specifically provide insights into how the performance in a pattern completion task at the network level depends on the length of the dendrites as well as on the overall size of the employed neurons.
Finally, we provide benchmarking results showcasing Arbor's runtime and memory efficiency compared to other simulators.

\section{Design and Implementation: Extensions of the Arbor core code}\label{sec:core}

In this section, we present the new extensions of the Arbor core code that are necessary for the implementation of synaptic plasticity models in Arbor. Examples of related models are presented in the following section \ref{sec:models}.

\subsection{Spike-time detection to simplify computation}\label{ssec:post_event_hook}

Many formulations of synaptic plasticity depend on the timing of pre- and postsynaptic spikes \citep{BiPoo1998,Song2000,GraupnerBrunel2012}. 
In neuroscience, spike-timing-dependent plasticity (STDP) serves as a valuable phenomenological model that can encapsulate the intricacies of synaptic plasticity with respect to its dependence on pre- and postsynaptic spike timing in a computationally efficient manner (cf. subsection~\ref{ssec:stdp}).
The simplicity of its formulation, focusing on the relative timing of spikes, makes STDP a practical choice for computational models that serve to understand learning processes in complex neural systems, without the necessity of a detailed molecular blueprint. 
Nevertheless, to capture the molecular and cellular mechanisms that underlie synaptic modification, more detailed models such as the calcium-based models used in subsections \ref{ssec:ca_plast_graupner_brunel}--\ref{ssec:stc_morpho_neuron_network} of this article are needed.

To enable the implementation of models with reduced complexity, such as STDP, we have introduced a hook named \verb+POST_EVENT+ that serves to detect spiking events in the postsynaptic neuron (e.g., somatic action potentials or dendritic spikes) and to transmit this information to another compartment or a synapse without explicit implementation of the physical transmission process (e.g., the backpropagation of action potentials). This is, inter alia, needed for STDP rules (cf. subsection \ref{ssec:stdp}) and calcium-based plasticity rules (cf. subsections \ref{ssec:ca_plast_graupner_brunel}--\ref{ssec:stc_morpho_neuron_network}).


\subsection{Multiple postsynaptic variables depending on pre- and postsynaptic spiking}\label{ssec:rr_halt}
The Arbor documentation \citep{arbor-docs-selection-policy} defines a selection policy as `Enumeration used for selecting an individual item from a group of items sharing the same label.' (where, for example, the items might be synapse objects with the label `exc\_input\_synapse'). 
Already present in previous Arbor versions, the \verb+round_robin+ policy enables to iterate over the items of an object group in a round-robin fashion, e.g., to iterate over synapses connecting to the same postsynaptic compartment.

We added the new selection policy \verb+round_robin_halt+, which enables to halt at the current item of the group until the \verb+round_robin+ policy is called again. This functionality is crucial to implement the independent update of multiple postsynaptic variables that depend on pre- and postsynaptic spiking. This is required, for instance, for large-scale network models including spike-driven postsynaptic calcium dynamics (see subsections \ref{ssec:stc_synapse_and_network} and \ref{ssec:stc_morpho_neuron_network}) that shall occur alongside the usual postsynaptic voltage dynamics. In such a case, the \verb+round_robin_halt+ policy serves to target both dynamics without having to define explicit labels for every individual connection in the network, which can save a tremendous amount of compute resources.


\subsection{Computation with stochastic differential equations}\label{ssec:stoch_diff_equations}

For numerous learning mechanisms, in particular also for some of the plasticity rules that are considered in the computational experiments presented here, random processes are required.
Such processes are often described by stochastic differential equations (SDEs). In general, the
coupled first-order equations for a vector of stochastic state variables $\textbf{X}$ can be expressed in their differential
form as

\begin{equation}
    d\textbf{X}(t) = \textbf{f}(t, \textbf{X}(t)) dt + \sum_{i=0}^{M-1} \textbf{g}_i(t,\textbf{X}(t)) d B_i(t),
\end{equation}

where the vector-valued function $\mathbf{f}$ denotes the deterministic differential, and the
last term represents the stochastic contribution. Here, the $M$ functions $\mathbf{g}_i$ with units
$[g_i] = [X]/\sqrt{t}$
are associated with standard Wiener processes $B_i$,
where $B_i(0) = 0$ almost surely, and $B_i(t) \sim \mathcal{N}(0, t)$ with units $[B_i]=\sqrt{t}$.

The stochastic integral is defined by It\^o's non-anticipative generalization of the Riemann–Stieltjes summation

\begin{equation}
    S_i = \int_{t_0}^{t_0+s} \textbf{g}_i(\tau,\textbf{X}(\tau)) d B_i(\tau) = \lim_{N\to\infty} \sum_{n=0}^{N-1} \textbf{g}_i(t_i,\textbf{X}(t_i)) \left( B(t_i) - B(t_{i-1})\right),
\end{equation}

where $t_0 < t_1 < \cdots < t_{N-1} = t_0 + s$, and $N \in \mathbb{N}$.
By introducing stationary Gaussian white noise $W_i$ such that $W_i(t) dt = dB_i(t)$,
the system of equations can be expressed using more common shorthand notation as
\begin{equation} \label{eq:SDE}
    \textbf{X}^\prime(t) = \textbf{f}(t, \textbf{X}(t)) + \sum_{i=0}^{M-1} \textbf{g}_i(t,\textbf{X}(t)) W_i(t).
\end{equation}

We equipped Arbor with the capability to numerically solve the system of SDEs depicted in Eq.~\ref{eq:SDE} using the Euler-Maruyama algorithm, a first-order stochastic Runge-Kutta method~\cite{kloeden1992}.
The algorithm for integrating the stochastic dynamics from discrete time step $t_{k}$ to $t_{k+1} = t_{k}+\Delta t$ comprises two steps:

\begin{enumerate}
    \item Drawing random variables $\Delta \mathbf{W} \sim \mathcal{N}(\mathbf{0}, \mathbf{Q}\Delta t)$, where $\mathbf{Q}$ denotes the
    correlation matrix of the white noises $W_i$,
    \item Computing $\hat{\mathbf{X}}(t_{k+1}) = \hat{\mathbf{X}}(t_k) + f(t_k, \hat{\mathbf{X}}(t_k)) \Delta t + \sum_{i=0}^{M-1} \mathbf{g}i(t_k,\hat{\mathbf{X}}(t_k)) \Delta W{i}$.
\end{enumerate}

Currently, we assume uncorrelated noise, $\mathbf{Q} = \mathbb{I}$.
Hence, to generate $M$ independent random samples for every instantiation of every stochastic process at each time step, a normally distributed noise source of sufficient quality is required.

Traditional pseudorandom number generators (PRNGs) prove inadequate for this context, as they typically generate a sequence of samples
through the evaluation of a recurrence relation $\varphi$ of order $k$. Here, the $n$th sample $u_n$ is contingent upon the $k$ preceding
values: $u_n = \varphi\left(n, u_{n-1}, u_{n-2}, \cdots, u_{n-k}\right)$.
For instance, the standard 64-bit implementation of the Mersenne-Twister algorithm in C\verb!++! necessitates the sequential updating of
a state comprising at least $19968$ bits ($k = 312$) to produce a series of independent random samples.

Consequently, we have opted for the utilization of counter-based PRNGs (CBPRNGs)~\cite{salmon2011}.
In CBPRNGs, each sample can be independently drawn by modulating the input to the generator function.
This input may be subdivided into a counter $c(n)$ and a key $\kappa(n)$, thus enabling the construction of a distinct input for every
required source of white noise, $u_{i, n} = \varphi\left(c_i(n), \kappa_i(n)\right)$. Owing to this characteristic and their stateless nature, CBPRNGs lend themselves well
to parallelization on both CPU and GPU architectures.

Specifically, we employ the \texttt{Threefry-4x64-12} algorithm from the \texttt{random123} library~\cite{random123}.
This algorithm’s input width, at $2 \times 256$ bits, affords ample capacity for uniquely encoding the white noise sources.
\texttt{Threefry-4x64-12} yields four independent, uniformly distributed values per invocation, which we cache for each
noise source and refresh only upon depletion.

To generate random numbers following a normal distribution, we employ the Box-Muller method,
ensuring uniform cache depletion across all noise sources as opposed to rejection-sampling based techniques.

In preceding versions of Arbor (before v0.8), the inclusion of random processes required modifying the C\verb!++! code
produced by Arbor's \texttt{modcc} compiler. This approach hindered the effective utilization of CBPRNGs and mandated the
manual crafting of solvers for SDEs. Presently, however, we have augmented
Arbor's NMODL dialect with a specialized solver method, denoted \verb!stochastic!, alongside a mechanism for
specifying independent noise sources via the \verb+WHITE_NOISE+ code block.
These enhancements enable the seamless handling of SDEs as described above.


\subsection{Diffusion of arbitrary particles}\label{ssec:diffusion_support}
 
Arbor's comprehension of neuronal morphology is built on the \emph{cable model} of neuronal dynamics:

\begin{equation}
    \frac{1}{C_\text{m}}\frac{\partial}{{\partial}t} U_\text{m} = \frac{\partial}{{\partial}x} \frac{1}{R_\text{l}} \frac{\partial}{{\partial}x} U_\text{m} + I_\text{m},
    \label{eq:cable}
\end{equation}

which describes the evolution of the membrane potential $U_\text{m}$ depending on time and one spatial dimension \cite{Rall1962,cable-equation-solution}. In this equation, $R_\text{l}$ denotes the axial (longitudinal) resistance and $C_\text{m}$ the membrane capacity. The current $I_\text{m}$ accounts for the radial transport of charges across the membrane via ion-channels (for more details, see Supplementary Fig.~S7).    
The term $\frac{\partial}{{\partial}x} \frac{1}{R_\text{l}} \frac{\partial}{{\partial}x} U_\text{m}$ describes a longitudinal current along the dendritic segment that results in charge equalization. 

It is usually assumed that this model is valid for a thin layer around the membrane where all changes to individually ionic concentrations -- commonly labeled $X_i$ and $X_o$ for the intra- and extracellular concentration of ion species $\mathrm{X}$ -- are equalized to that of a surrounding internal or external buffer solution. 
This buffering is modeled as an infinitely fast process, such that any alterations are visible only on timescales of the numerical timestep.
The trans-membrane current $I_\text{m}$ can be expressed as a function of individual ion species $\mathrm{X}$:
\begin{align*}
    I_\text{m} &= \sum\limits_{\mathrm{X}} g_\mathrm{X}\cdot(U_\text{m} - E_\mathrm{X}),
\end{align*}
with
\begin{align*}
    E_\mathrm{X} &= \frac{RT}{z_\mathrm{X}F}\cdot\ln\left(\frac{X_o}{X_i}\right),
\end{align*}
where the ion channel models produce the conductivities $g_\mathrm{X}$ and the reversal potentials $E_\mathrm{X}$, with universal gas constant $R$, Faraday constant $F$, temperature $T$, and charge number $z_\mathrm{X}$.

Note that models of neuronal dynamics that include the resolution of individual ions in the evolution of the membrane potential are tractable but computationally more demanding than the cable model \cite{saetra2020}.
The diffusion of particles along dendrites is, nevertheless, a critical element for many computational neuron models.
As mentioned before, a rigorous model for the transport of ions is feasible, but involves a different equation for charge equalization as opposed to Eq.~\ref{eq:cable}. Namely, it requires handling the changes in the intra- and extracellular concentration of particles through molar fluxes (including the buffering because now the associated timescale has become relevant), and -- closing the feedback loop -- the Nernst equation for computing the individual reversal potentials.
The alternative of mixing models would lead to a flawed formulation, where particles are transported by diffusion despite already being included in the longitudinal currents of Eq.~\ref{eq:cable}.
Thus, we decided to implement diffusion of arbitrary particles in Arbor as if the relevant species were strictly neutral, i.e. no additional flow of charges is considered, neither along the dendrite nor across the membrane.
The physical model for diffusion of the concentration $X$ of the specific particle species is then simply given by

\begin{equation}
  \frac{\partial}{{\partial}t} X = \frac{\partial}{{\partial}x} D \frac{\partial}{{\partial}x} X + \phi,
  \label{eq:diffusion}
\end{equation}

where we have the diffusivity $D$ and the molar flux $\phi$ across the membrane, from, or to internal stores.
This equation is in shape identical to the cable equation, which allows us to leverage Arbor's existing, highly optimized solver. 
The diffusive model is decoupled from the cable model and exposed via a separate quantity $X_d$ to NMODL.
Users can --- if desired --- retrofit the interaction with the cable model by assigning the appropriate mechanisms formulated in NMODL:

\begin{align*}
    I_\text{m} = \sum\limits_{\mathrm{X}} I_\mathrm{X} &= \sum\limits_{\mathrm{X}} g_\mathrm{X}\cdot(U - E_\mathrm{X}),\\
    \frac{\partial}{\partial t} X_d &= \phi_{\mathrm{X}} + \frac{I_\mathrm{X}}{z_\mathrm{X}F}, \\
    E_\mathrm{X} &= \frac{RT}{z_\mathrm{X}F}\cdot\ln\left(\frac{X_o}{X_d}\right),
\end{align*}

yielding a closed model together with Eqs. \eqref{eq:cable} and \eqref{eq:diffusion}, however, at the cost of the inherent issues described above.


\section{Results: Computational modeling with synaptic plasticity}\label{sec:models}

In this section, we present representative models of synaptic plasticity that we implemented in Arbor, serving as examples to demonstrate the full functionality of the newly implemented features. 
As we have made the respective code freely available, readers can simply reuse or adapt the models for their own investigations with Arbor. For references to the code bases used for the specific simulations, please see the data availability statement at the end of this article. 

Furthermore, we provide benchmarking results to compare the runtime and memory performance of the simulation of different network models in Arbor and other simulators.

Please note that since the implementation of plasticity dynamics is the main target of this paper, we only present the model equations that correspond to those. For parts of the considered models other than the plasticity dynamics, please refer to the cited literature and provided code.

\begin{figure}
  \includegraphics[width=\textwidth]{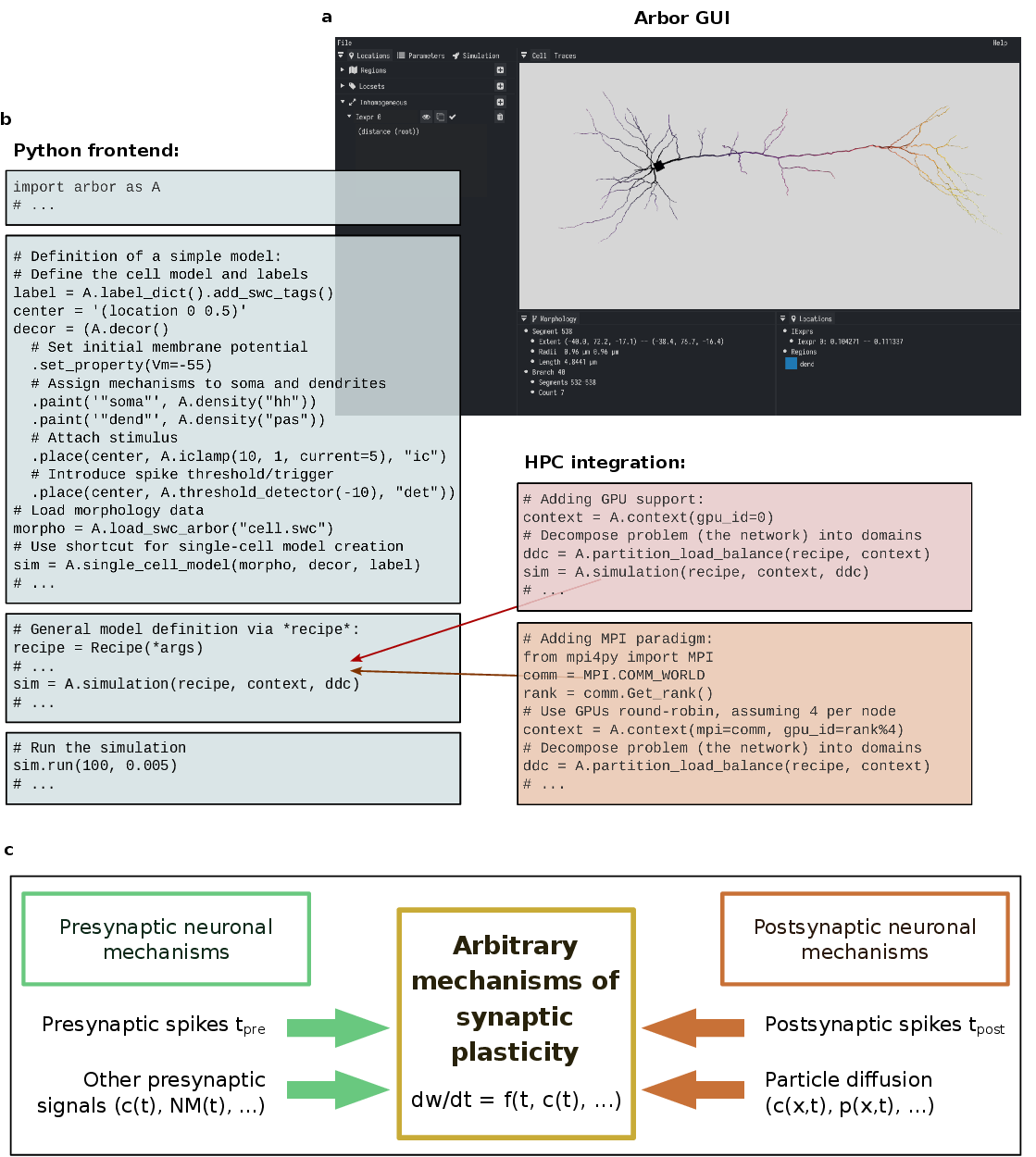}
  \caption{\small \textbf{Overview of the extended Arbor framework.} \textbf{(a)} Graphical presentation of cell morphologies via \texttt{Arbor~GUI} (v0.8.0-dev-065b1c9 shown here). \textbf{(b)} Key code features for model simulation with Arbor. On the left: facilitated definition of models via the Python frontend by setting up a so-called \emph{recipe} (where in some cases, there are shortcuts that further accelerate the definition of certain simple models). On the right: modular integration of high performance computing (HPC) hardware such as graphics processing units (GPUs) and message passing interface (MPI) ranks. \textbf{(c)} General approach of the new plasticity framework, enabling to simulate a wide variety of synaptic plasticity mechanisms in arbitrary model paradigms (examples of which are provided in section \ref{sec:models}).}
  \label{fig:ui}
\end{figure}

Fig.~\ref{fig:ui}a shows an overview of essential features of the Arbor user interface, which can be used to implement and simulate models of plastic neuronal networks. Fig.~\ref{fig:ui}b summarizes the main contribution of our present work: a simulation framework that enables to target arbitrary mechanisms of synaptic plasticity.

\subsection{Spike-timing-dependent plasticity}\label{ssec:stdp}

Spike-timing-dependent plasticity (STDP) is a phenomenon that is reported in a number of experimental studies \citep{Gerstner1996,MageeJohnston1997,Markram1997,BiPoo1998} and described by various theoretical models (cf. \cite{Morrison2008}). 
The earlier models of STDP have provided a relatively simple form of synaptic plasticity that only depends on \emph{the specific timing of pre- and postsynaptic spikes} via the temporal differences $t^n_\textrm{pre} - t^m_\textrm{post}$ ($m,n \in \mathbb{N}$) between pre- and postsynaptic spikes. In this case, plasticity does not directly depend on the spike rate. As a first step, we implemented a commonly used description \citep{Song2000} given by:

\begin{align}
\frac{da_\textrm{pre}(t)}{dt} & = -\frac{a_\textrm{pre}(t)}{\tau_\textrm{pre}} + A_\textrm{pre} \cdot \sum \limits_n \delta (t - t^n_\textrm{pre}),\label{eq:stdp_ode_1}\\
\frac{da_\textrm{post}(t)}{dt} & = -\frac{a_\textrm{post}(t)}{\tau_\textrm{post}} + A_\textrm{post} \cdot \sum \limits_m \delta (t - t^m_\textrm{post}),\label{eq:stdp_ode_2}\\
\frac{dw(t)}{dt} & = a_\textrm{pre}(t) \cdot \sum \limits_m \delta (t - t^m_\textrm{post}) + a_\textrm{post}(t) \cdot \sum \limits_n \delta (t - t^n_\textrm{pre}).\label{eq:stdp_ode_3}
\end{align}

Here, the constants $\tau_\textrm{pre}$ and $\tau_\textrm{post}$ describe the decay of the eligibility traces $a_\textrm{pre}(t)$ and $ a_\textrm{post}(t)$ induced by pre- and postsynaptic spikes, and $\delta(\cdot)$ represents the Dirac delta distribution. The amplitudes $A_\textrm{pre} > 0\unit{\textmu{S}}$ and $A_\textrm{post} < 0\unit{\textmu{S}}$ define the strengthening and weakening of the synaptic weight $w(t)$ that follow the occurrence of spikes, which takes more effect the closer together pre- and postsynaptic spikes occur in time. Note that the synaptic weight $w(t)$ is updated for each complete pair of pre- and postsynaptic spikes. The synaptic weight is initialized at the baseline value $w_0$, and its contribution to postsynaptic potentials is clipped at a value $w_\textrm{max}$. For the parameter values, see Table~\ref{tab:parameters_model_stdp}.

To simulate this model with Arbor, we implemented a single, conductance-based excitatory synapse. This synapse was connected to a Leaky Integrate-and-Fire (LIF) neuron and was stimulated with Poissonian spike input (see Fig.~\ref{fig:stdp_homeostasis_calcium}a). In addition, an inhibitory synapse was introduced to stabilize the dynamics, also driven by Poissonian spike input. Both the excitatory synapse and the LIF neuron were implemented as custom mechanisms in Arbor (defined via NMODL scripts). Note that for this implementation, the \verb+POST_EVENT+ hook, which we added to Arbor's NMODL dialect, is needed (cf. subsection \ref{ssec:post_event_hook}) -- the hook is called whenever a threshold detector on a cell is triggered. In the present case, this is when the postsynaptic neuron spikes. 

We evaluated the functionality of the STDP implementation by comparing between Arbor and the Brian 2 simulator, which provides a suitable cross-validation due to the different implementations of the numerical solver as well as the algorithmic data representation. We observe a good match between the presented measures recorded in Brian 2 and Arbor (Fig.~\ref{fig:stdp_homeostasis_calcium}c and Supplementary Fig. S1a--c), which we quantified via the values of coefficient of variation (CV) and root mean square error (RMSE), computed using \texttt{scikit-learn} in version 1.3.2. Regarding the spike times, we also encounter only minimal differences, which we quantified by computing the spike time mismatch (detailed in Supplementary Fig. S1d).

In addition to this experiment, we simulated a range of time delay settings to obtain a classical STDP time window, which also shows a very good match with Brian 2 simulations (Fig.~\ref{fig:stdp_homeostasis_calcium}d) as well as with the theoretical expectation (Supplementary Fig.~S1e).

\begin{table}[!ht]
\small
\begin{center}
\begin{tabular}{|l|p{2.8cm}|p{5.7cm}|p{2.5cm}|}
    \hline {Symbol} & {Value} & {Description} & {Refs.} \\
	\hline
    \hline {$w_0$} & {$1.0\unit{\textmu{S}}$} & {Baseline of the synaptic weight} & {This study} \\
    \hline {$\tau_\textrm{pre}$} & {$20.0\unit{ms}$} & {Decay of the eligibility trace of presynaptic spikes} & {\citep{Song2000,brian-tutorial-intro-2}} \\
    \hline {$\tau_\textrm{post}$} & {$\{10.0, 20.0\}\unit{ms}$} & {Decay of the eligibility trace of postsynaptic spikes} & {\citep{Song2000,brian-tutorial-intro-2}} \\
    \hline {$A_\textrm{pre}$} & {$\{0.3, 0.01\}\unit{\textmu{S}}$} & {Strengthening amplitude} & {\citep{Song2000,brian-tutorial-intro-2}} \\
    \hline {$A_\textrm{post}$} & {$\{-0.2,-0.0105\}\unit{\textmu{S}}$} & {Weakening amplitude} & {\citep{Song2000,brian-tutorial-intro-2}} \\
    \hline {$w_\textrm{max}$} & {$10.0\unit{\textmu{S}}$} & {Maximum synaptic weight contributing to postsynaptic potentials} & {\citep{brian-tutorial-intro-2}} \\
	\hline
\end{tabular}
\caption{\small Parameters for the plain STDP model. In the case that two values are given, the first value was used for the detailed analysis shown in Fig.~\ref{fig:stdp_homeostasis_calcium}c and Supplementary Fig. S1a--d, and the second value was used to obtain the classical curve shown in Fig.~\ref{fig:stdp_homeostasis_calcium}d (also cf.~\citep{brian-tutorial-intro-2}).\label{tab:parameters_model_stdp}}
\end{center}
\end{table}

\clearpage

\begin{figure}
    \captionsetup{labelformat=original}
    \centering
    \includegraphics[width=\textwidth]{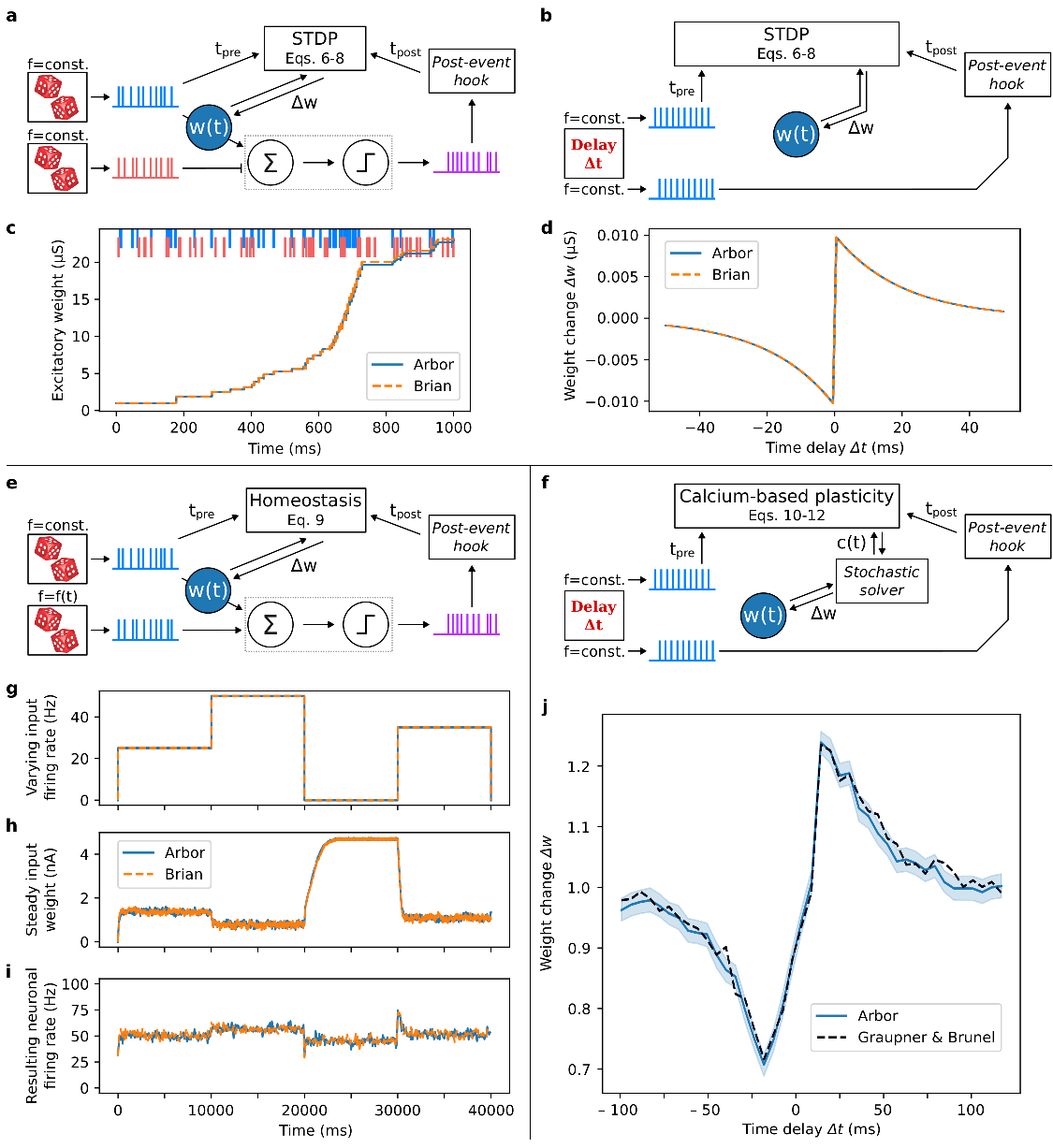}
    \caption{\small \textbf{Classical spike-timing dependent plasticity (STDP), spike-driven homeostasis, and calcium-driven synaptic plasticity in Arbor.} Arbor implementations (in lighter blue) are cross-validated by comparison to Brian 2 (in orange) or a custom simulator. New features of the Arbor core code are highlighted in italic. \textbf{(a)} STDP paradigm where two Poisson spike sources stimulate an excitatory and an inhibitory synapse connecting to a single neuron (spikes shown in blue and red, respectively). The excitatory connection undergoes STDP (results shown in (c)). Image of dice from Karen Arnold/publicdomainpictures.net. \textbf{(b)} STDP paradigm where two regular spike trains, phase-shifted by delay $\Delta{t}$, drive the weight dynamics of a single synapse (results shown in (d)). \textbf{(c)} Strength of the excitatory synapse, subject to STDP, as shown in (a) (goodness of fit between Arbor and Brian 2: $\mathrm{CV} > 0.999$, $\mathrm{RMSE} = 1.064\unit{{\textmu}S}$). \textbf{(d)} Classical STDP curve, obtained as detailed in (b) ($\mathrm{CV} > 0.999$, $\mathrm{RMSE} = 0.001\unit{{\textmu}S}$). \textbf{(e)} Homeostasis with two Poisson spike sources connected to an LIF neuron via current-based delta synapses (results shown in (g--i), averaged over $50$ trials). One of these Poisson inputs spikes at a fixed rate and is plastic, while the other spikes at a varying rate and is static.}
    \label{fig:stdp_homeostasis_calcium}
\end{figure}
\clearpage
\begin{figure}
    \captionsetup{labelformat=adja-page}
    \ContinuedFloat
    \caption{\small \textbf{(f)} Paradigm of calcium-based, spike-timing- and rate-dependent synaptic plasticity, using the model by Graupner \& Brunel \citep{GraupnerBrunel2012}. Two regular spike trains, phase-shifted by delay $\Delta{t}$, drive the stochastic weight dynamics of a single synapse (results shown in (j)). \textbf{(g)} Time course of the varying rate of the input in (e). \textbf{(h)} Strength of the plastic synapse, subject to the dynamics given in (e) ($\mathrm{CV} = 0.996$, $\mathrm{RMSE} = 0.307\unit{nA}$). \textbf{(i)} Firing rate of the neuron shown in (e) in the presence of homeostatic plasticity dynamics ($\mathrm{CV} = 0.508$, $\mathrm{RMSE} = 1.981\unit{Hz}$). \textbf{(j)} Calcium-driven synaptic plasticity as shown in (f). Reproduction of the numerical DP curve from Figure 2 of the related paper \citep{GraupnerBrunel2012} (the mean is given by the dark dashed line). 
    Every synapse is subject to $60$ spike pairs presented at $1\unit{Hz}$. 
    Arbor results were averaged over $4000$ trials, the solid blue line indicates the mean and the shaded region the $95\%$ confidence interval.
    Quantification of deviation between the mean curves: $\mathrm{CV} = 0.987$, $\mathrm{RMSE} = 0.123$. Note that the generation of this plot is now also demonstrated in an Arbor tutorial \citep{arbor-tutorial-calcium-model}.}
\end{figure}


\subsection{Spike-driven homeostatic plasticity}\label{ssec:homeostasis}

After implementing an STDP rule at a single synapse,
we now consider a type of synaptic plasticity that depends on \emph{multiple presynaptic stimuli onto the same postsynaptic neuron}.

Spike-based homeostatic plasticity describes the finding that the strength of a synapse adapts according to the spiking activity of the postsynaptic neuron -- hence, the synaptic strength is up- or downregulated to maintain a certain activity of the neuron (cf.~\citep{ZenkeGerstner2017}). 
By using the newly implemented \verb+POST_EVENT+ hook (introduced in subsection \ref{ssec:post_event_hook}), we could employ Arbor to simulate spike-driven homeostasis. 
To show this, we connect (similar to \citep{Breitwieser2015}) an LIF neuron to two Poisson-stimulus inputs -- one with a varying spike rate and the other with a fixed spike rate. The weight of the synapse for the varying-rate input is kept static, while the weight of the synapse for the fixed-rate input is plastic. The plasticity of the latter synapse should cause the neuron, in a homeostatic manner, to maintain a firing rate that is determined by the parameters of the plasticity rule. This is realized by the following weight dynamics for the plastic synapse:

\begin{align}
\frac{dw(t)}{dt} & = \Delta{w}_{+} \cdot \sum \limits_n \delta (t - t^n_\textrm{pre}) + \Delta{w}_{-} \cdot \sum \limits_m \delta (t - t^m_\textrm{post}),\label{eq:homeostatic_plasticity}
\end{align}
where the constants $\Delta{w}_{+}$ and $\Delta{w}_{-}$ describe the weight changes upon the occurrence of pre- and postsynaptic spikes at times $t^n_\textrm{pre}$ and $t^m_\textrm{post}$ ($n,m \in \mathbb{N}$). The weight is initialized at value $w_\text{init}$ (see Table~\ref{tab:parameters_model_homestasis}).

The resulting weight and firing rate dynamics, along with the varying input rate, are shown in Fig.~\ref{fig:stdp_homeostasis_calcium}g-i and Supplementary Fig.~S1f. We can see that in the case with homeostasis, the resulting firing rate is maintained at values around $50\unit{Hz}$ (Fig.~\ref{fig:stdp_homeostasis_calcium}i), while in the case without homeostasis (Supplementary Fig.~S1f), the resulting firing rate is mainly imposed by the input rate. Note that for the time period with input rate $0$ (from $t=20\unit{s}$ to  $30\unit{s}$), the homeostatic weight adjustment can only happen to a limited extent since we do not allow the weights to increase beyond ${w}_\text{max}$. 
We cross-validated our Arbor implementation with Brian 2, drawing from an existing example implementation \citep{brian-homeostasis-example}.

\begin{table}[!ht]
\small
\begin{center}
\begin{tabular}{|l|p{2.5cm}|p{6.0cm}|p{2.5cm}|}
    \hline {Symbol} & {Value} & {Description} & {Refs.} \\
	\hline
	\hline {$w_\text{init}$} & {$0.00\unit{nA}$} & {Baseline weight of the plastic (fixed-rate input) synapse} & {\citep{Breitwieser2015}}\\
	\hline {${w}_\text{max}$} & {$5.00\unit{nA}$} & {Maximum weight of the plastic (fixed-rate input) synapse} & {\citep{Breitwieser2015}}\\
	\hline {$\Delta{w}_{+}$} & {$0.35\unit{nA}$} & {Weight change due to presynaptic spike} & {\citep{Breitwieser2015}}\\
	\hline {$\Delta{w}_{-}$} & {$-0.35\unit{nA}$} & {Weight change due to postsynaptic spike} & {\citep{Breitwieser2015}}\\
	\hline {$w_\textrm{varying}$} & {$3.50\unit{nA}$} & {Weight of the varying-rate input synapse} & {\citep{Breitwieser2015}}\\
	\hline
\end{tabular}
\caption{\small Parameters for the homeostatic plasticity model.\label{tab:parameters_model_homestasis}}
\end{center}
\end{table}


\subsection{Calcium-based synaptic plasticity}\label{ssec:ca_plast_graupner_brunel}

Following the implementation of a simple phenomenological STDP rule (subsection \ref{ssec:stdp}) as well as spike-driven homeostatic plasticity (subsection \ref{ssec:homeostasis}), we now consider a more complex rule of synaptic plasticity that requires multiple postsynaptic variables. This rule describes the potentiation and depression of synaptic strength depending on \emph{the postsynaptic calcium concentration}, which is driven by pre- and postsynaptic spiking activity. It was presented by Graupner and Brunel in 2012 \citep{GraupnerBrunel2012} and has since been used widely. In comparison to the phenomenological STDP rule, the calcium-based rule adapts the synaptic weight dynamics such that they depend on spike timing and spike rates, both of which have been shown to be important features of long-term synaptic plasticity \citep{Sjoestroem2001}. 

The change of the synaptic weight in this model is given by the following equation:

\begin{align}
\tau_w
\frac{dw_{ji}(t)}{dt} = & - w_{ji}(t) \cdot \left(1-w_{ji}(t)\right)  \cdot \left(w_{*}-w_{ji}(t)\right)\nonumber\\
& + \gamma_\textrm{p} \cdot \left(1-w_{ji}(t)\right)\cdot\Theta\left[c_{ji}(t)-\theta_\textrm{p}\right]\nonumber\\
& - \gamma_\textrm{d} \cdot w_{ji}(t) \cdot\Theta\left[c_{ji}(t)-\theta_\textrm{d}\right]\nonumber\\
& + \xi(t),\label{eq:weight_graupner_brunel}
\end{align}
where $\tau_\textrm{w}$ is a time constant, $w_{*}$ defines the boundary between the basins of attraction for potentiation and depression, $\gamma_\textrm{p}$ and $\gamma_\textrm{d}$ are the potentiation and depression rates, $c_{ji}(t)$ is the calcium concentration at the postsynaptic site, and $\theta_\textrm{p}$ and $\theta_\textrm{d}$ are thresholds for triggering potentiation and depression, respectively. 
Moreover, $\Theta[\cdot]$ denotes the Heaviside theta function, and 

\begin{align}
\xi(t) = \sqrt{\tau_w \left(\Theta\left[c(t)-\theta_\textrm{p}\right] + \Theta\left[c(t)-\theta_\textrm{d}\right]\right)}\cdot\sigma_\textrm{pl}\cdot\Gamma(t) \label{eq:plasticity_noise_1}
\end{align}
is a noise term with scaling factor $\sigma_\textrm{pl}$ and Gaussian white noise $\Gamma(t)$, which has a mean value of zero and a variance of $1/dt$ (cf. \citep{Gillespie1996}). Note that to implement the noise term, support for stochastic differential equations was needed, which we have added to the Arbor core code as described above in subsection \ref{ssec:stoch_diff_equations}.

Finally, the dynamics of the calcium concentration is given by the following equation:
\begin{equation}
\frac{dc(t)}{dt} = -\frac{c(t)}{\tau_\textrm{c}} + c_\textrm{pre}\cdot\sum\limits_{n}\delta(t-t^n_\textrm{pre}-t_\textrm{c,delay}) + c_\textrm{post}\cdot\sum\limits_{m}\delta(t-t^m_\textrm{post}),\label{eq:ca_model_graupner_brunel}
\end{equation}
where $\tau_c$ is a time constant, $c_\textrm{pre}$ and $c_\textrm{post}$ are increases in the intracellular calcium concentration of the dendritic spine induced by pre- and postsynaptic spikes at times $t^n_\textrm{pre}$ and $t^m_\textrm{post}$, $t_\textrm{c,delay}$ is the delay of the presynaptic contributions, and $\delta(\cdot)$ is the Dirac delta distribution.

Fig. \ref{fig:stdp_homeostasis_calcium}j shows the results of our Arbor implementation for the weight change over the delay between pre-and postsynaptic spikes (analogously to Fig.~\ref{fig:stdp_homeostasis_calcium}d), cross-validated with the numerical results by the original study \citep{GraupnerBrunel2012}.

\begin{table}[!ht]
\small
\begin{center}
\begin{tabular}{|l|p{2.5cm}|p{6.0cm}|p{2.5cm}|}
    \hline {Symbol} & {Value} & {Description} & {Refs.} \\
    \hline
    \hline {$w_0$} & {$0.0$ or $1.0$} & {Initial value of the synaptic weight (drawn randomly)} & {This study}\\
	\hline {$w_{*}$} & {$0.5$} & {Boundary between the basins of attraction for potentiation and depression} & {\citep{GraupnerBrunel2012}}\\
	\hline{$t_\textrm{c,delay}$} & {$13.7\unit{ms}$} & {Delay of postsynaptic calcium influx after presynaptic spike} & {\cite{GraupnerBrunel2012}}\\
	\hline {$c_\textrm{pre}$} & {$1$} & {Presynaptic calcium contribution, in vivo adjusted} & {\cite{GraupnerBrunel2012}}\\
	\hline {$c_\textrm{post}$} & {$2$} & {Postsynaptic calcium contribution, in vivo adjusted} & {\cite{GraupnerBrunel2012}}\\
	\hline {$\tau_c$} & {$20\unit{ms}$} & {Calcium time constant} & {\cite{GraupnerBrunel2012}}\\
	\hline {$\tau_w$} & {$150\unit{s}$} & {Weight dynamics time constant} & {\cite{GraupnerBrunel2012}}\\
	\hline {$\gamma_\textrm{p}$} & {$321.808$} & {Potentiation rate} & {\cite{GraupnerBrunel2012}}\\
	\hline {$\gamma_\textrm{d}$} & {$200$} & {Depression rate} & {\cite{GraupnerBrunel2012}}\\
	\hline {$\theta_\textrm{p}$} & {$1.3$} & {Calcium threshold for potentiation} & {\cite{GraupnerBrunel2012}}\\
	\hline {$\theta_\textrm{d}$} & {$1$} & {Calcium threshold for depression} & {\cite{GraupnerBrunel2012}}\\
	\hline {$\sigma_\textrm{pl}$} & {$2.8248$} & {Standard deviation for plasticity fluctuations} & {\cite{GraupnerBrunel2012}}\\
	\hline
\end{tabular}
\caption{\small Parameters for the implementation of the calcium-based plasticity model by Graupner \& Brunel \cite{GraupnerBrunel2012}. Note that as in the original mathematical model, the weights are kept without physical unit.\label{tab:parameters_model_graupner-brunel}}
\end{center}
\end{table}


\subsection{Heterosynaptic calcium-based plasticity in dendrites}\label{ssec:heterosyn_ca_plast}

As a next step, we use a calcium-based plasticity rule slightly different to the one in the previous subsection, with the aim to simulate the \emph{spread of calcium in a dendritic branch}, which enables us to model \emph{heterosynaptic plasticity}. This model serves as an example of our diffusion extension for the Arbor core code, which has been described in subsection \ref{ssec:diffusion_support}.

Homosynaptic plasticity and heterosynaptic plasticity are two forms of plasticity that play crucial roles in shaping neural connections. Homosynaptic plasticity involves changes within a specific neural pathway or synapse in response to repeated stimulation or learning, leading to the strengthening or weakening of that connection depending on factors such as frequency or duration of stimulation.
On the other hand, heterosynaptic plasticity is a broader phenomenon where stimulation of one synapse induces changes in other, unstimulated synapses. 
Here, we consider a calcium-based heterosynaptic plasticity rule \citep{HirataniFukai2017} that is based on observations at the level of a single neuron \cite{GraupnerBrunel2012,Lee2016}. 

Dendrites are crucial components for the information processing in neurons, as they receive signals from other neurons and integrate them to generate a particular response. Spiny structures on the dendrites can serve to receive synaptic inputs and at the same time undergo plastic changes \cite{DendritesBook}. Here, we consider a model that describes a number of such spines on a single dendritic branch \citep{Shafiee2024}.
The state and strength of these spines are subject to the previously mentioned calcium-based plasticity rule \citep{HirataniFukai2017}. 
We describe the synaptic input to a specific spine via

\begin{equation}
\begin{aligned}
I^\text{Ca}_\textrm{spine}(x,t_i) = \sum_{i}I_0\cdot e^{-(t-t_i)/\tau_I}\cdot\Theta(t-t_i),
\end{aligned}
\label{eq:ca_injection_current_spine}
\end{equation}

which induces an elevated level of calcium at the target spine via monoexponential contributions with amplitude $I_0$ and time constant $\tau_I$. However, due to the calcium diffusion dynamics in our system, other unstimulated and inactive spines will also experience changes in their calcium level. These changes depend on the spine location with respect to the stimulated spine(s) as well as the temporal characteristics of the stimulation (i.e., frequency, duration, and delay). In our simulations, pre-synaptic spike events arrive at active spines at times $t_i$ as a regular spike train with a time interval of $10\unit{ms}$. Note that we inject synaptic input at the top point of the spine heads, so the term $I^\text{Ca}_\textrm{spine}(x,t_i)$ is zero for all other regions (in particular, the dendritic shaft). The calcium diffusion in the dendritic branch and in the spines is then described by the following equation:

\begin{equation}
\begin{aligned}
& \frac{\partial C(x,t)}{\partial t}=D\frac{\partial^2 C(x,t)}{\partial x^2} -\frac{C(x,t)}{\tau_C}+ w_i \cdot I^\text{Ca}_\textrm{spine}(x,t_i), \\  
\end{aligned}
\label{eq:ca_diffusion_dendrite}
\end{equation}

where $C(x,t)$ is the calcium concentration, $D$ is the diffusion constant, and $\tau_C$ is the calcium decay time constant. 
The synaptic strength is computed as outlined below:

\begin{align}
\frac{dw_i(t)}{dt} = \, & (1 - w_i(t)) \cdot  \gamma_{p} \cdot \Theta\left[C(t)-\theta_{p}\right]\nonumber\\
& - w_i(t) \cdot \gamma_{d} \cdot \Theta\left[C(t)-\theta_{d}\right],\label{eq:heterosyn_weight}
\end{align}
where $w_i$ is the synaptic weight of spine $i$, the constants $\gamma_{p}$ and $\gamma_{d}$ quantify the rate of synaptic strengthening or weakening during potentiation and depression, and $\theta_{p}$ and $\theta_{d}$ are the calcium threshold values for triggering potentiation and depression, respectively.

\begin{figure}[!ht]
    \centering
    \includegraphics[width=\textwidth]{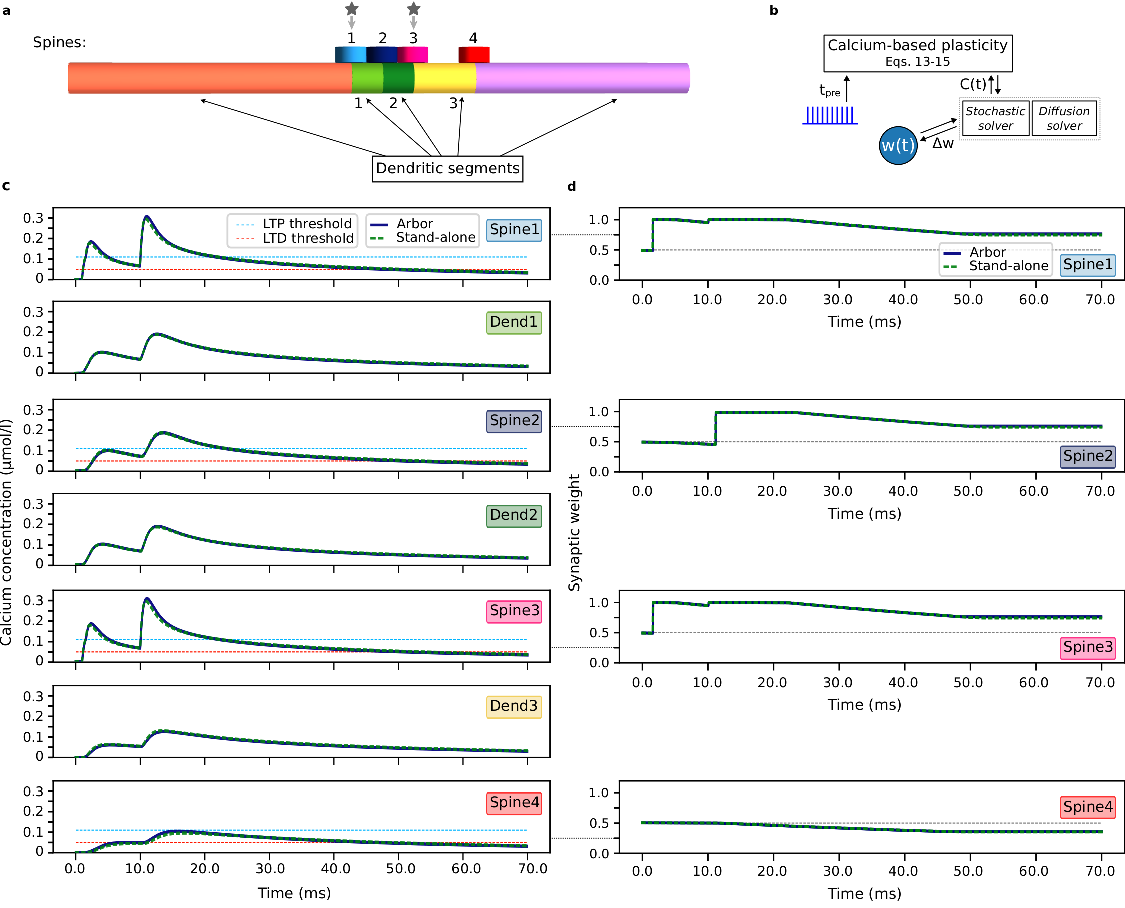}
    \caption{\small \textbf{Calcium-driven heterosynaptic plasticity in four spines on a dendritic branch.} Calcium is first introduced in spines $1$ and $3$ through synaptic input. Subsequently, calcium spatially distributes across the dendrite (according to Eq.~\ref{eq:ca_diffusion_dendrite}), which influences synaptic plasticity at other synapses, promoting either depression or potentiation. The parameter values are provided in Table~\ref{tab:parameters_model_heterosyn_calcium}. \textbf{(a)} Illustration generated using \texttt{Arbor~GUI} \citep{Arbor-GUI-0-8-0}. Each segment is represented by a different color, and a segment can consist of multiple compartments. Spines 1--4 are located at $x=-1.0,0.0, 1.0,3.0 \unit{{\textmu}m}$, respectively. For the purpose of visualization, the morphology has been clipped at $-10\unit{{\textmu}m}$ and $+10\unit{{\textmu}m}$, before scaling the dendrite along the x-axis by 2. \textbf{(b)} Paradigm of synaptic plasticity that depends on a spike-timing- and rate-dependent, diffusive calcium concentration (cf.~\citep{Shafiee2024}). Regular spike trains induce calcium injection in specific spines, eventually leading to weight changes (results shown in (c--d)). New features of the Arbor core code are highlighted in italic. \textbf{(c)} The change in the calcium level of each spine, and of the dendritic segments in between, in response to the stimulation to spines $1$ and $3$. Quantification of deviation between the simulators given by ($\mathrm{CV}, \mathrm{RMSE}$):  spine 1: ($0.991$, $0.005 \unit{{\textmu}mol/l}$); spine 2: ($0.998$, $0.002 \unit{{\textmu}mol/l}$); spine 3: ($0.989$, $0.005 \unit{{\textmu}mol/l}$); spine 4: ($0.967$, $0.004 \unit{{\textmu}mol/l}$; dendrite location 1: ($0.998$, $0.002 \unit{{\textmu}mol/l}$); dendrite location 2: ($0.998$, $0.002 \unit{{\textmu}mol/l}$); dendrite location 3: ($0.998$, $0.003 \unit{{\textmu}mol/l}$)). \textbf{(d)} Synaptic weight changes, which follow the calcium level of the spines. Spines $1$--$3$ undergo synaptic potentiation (elevated synaptic weights), while spine $4$ undergoes depression (reduced synaptic weight). Quantification of deviation between the simulators ($\mathrm{CV}, \mathrm{RMSE}$): spine 1: ($0.982$, $0.015$); spine 2: ($0.993$, $0.014$); spine 3: ($0.981$, $0.016$); spine 4: ($0.996$, $0.003$).
    \label{fig:heterosynaptic_plasticity_dendrite}}
\end{figure}

The dendritic branch in our model includes four spines, each with head segments, as shown in Fig.~\ref{fig:heterosynaptic_plasticity_dendrite}a. Synaptic inputs are applied to spines $1$ and $3$, which increases the level of calcium in these spines. Next, by means of diffusion, the level of calcium in spines $2$ and $4$ increases as well (Fig.~\ref{fig:heterosynaptic_plasticity_dendrite}c). This causes the synaptic weight at these spines to change in a manner that depends on the spatial proximity to the active spines. Fig.~\ref{fig:heterosynaptic_plasticity_dendrite}d shows that spine $2$ undergoes heterosynaptic potentiation as a result of its proximity to the active spines $1$ and $3$, whereas spine $4$ undergoes heterosynaptic depression due to its remote location and consequently lower levels of calcium. The Arbor results are cross-validated with a custom stand-alone simulator written in Python \citep{HeterosynapticCalciumCode}.

\begin{table}[!ht]
\small
\begin{center}
\begin{tabular}{|l|p{2.5cm}|p{6.0cm}|p{2.5cm}|}
    \hline 
    {Symbol} & {Value} & {Description} & {Refs.} \\
    \hline
    \hline
    $\gamma_{p}$ & $90$ & Potentiation rate & This study\\
    \hline
    $\gamma_{d}$ & $0.01$ & Depression rate & This study\\
    \hline
    $\theta_{p}$ & $0.11\unit{{\textmu}mol/l}$ & Calcium threshold for potentiation & This study\\
    \hline
    $\theta_{d}$ & $0.05\unit{{\textmu}mol/l}$ & Calcium threshold for depression & This study\\
    \hline
    $r_\textrm{head}$ & $1.0\unit{{\textmu}m}$ & Spine head radius & {\cite{Araya2014}}\\
    \hline
    $l_\textrm{head}$ & $1.0\unit{{\textmu}m}$ & Spine head length & {\cite{Araya2014}}\\
    \hline
    $r_\textrm{dendrite}$ & $1.0\unit{{\textmu}m}$ & Dendrite radius & {\cite{Araya2014}}\\
    \hline
    $l_\textrm{dendrite}$ & $80.0\unit{{\textmu}m}$ & Dendrite length & {This study}\\
    \hline
    {$\Delta{l}_\textrm{comp}$} & {$1.0\unit{\textmu{m}}$} & {Length of one compartment} & {This study}\\
    \hline
    $\tau_C$ & $100\unit{ms}$ & Calcium decay time constant & {\cite{Yasuda2017}}\\
    \hline
    $\tau_I$ & $1\unit{ms}$ & Injection current time constant & {\citep{Hausser1997,Gerstner2014}}\\
    \hline
    $\gamma$ & 0.11  & Fraction of current carried by $\text{Ca}^{\text{2+}}$ & {\cite{Biess2011}}\\
    \hline
    $I_0$ & $4.0\unit{pA }$ & Injection current (Arbor implementation) & {\cite{Biess2011}}\\
    \hline
    $I_0$ & $5.5\unit{pA }$ & Injection current (stand-alone implementation) & {\cite{Biess2011}}\\
    \hline
    $D$ & $2.2\cdot10^{-10}\unit{m\textsuperscript{2}/s}$ & Calcium diffusion constant & {\cite{Allbritton1992,Means2006,Biess2011}}\\
    \hline
\end{tabular}
\caption{\small Parameters for the calcium-based heterosynaptic plasticity model (also cf. Fig.~\ref{fig:heterosynaptic_plasticity_dendrite}).
Note that the injection current amplitude $I_0$ varies across implementations due to the differences mentioned in the main text.
\label{tab:parameters_model_heterosyn_calcium}}
\end{center}
\end{table}

Although the results from both simulators match very well, there is a specific difference between the models that should be mentioned. Namely, the two implementations use different models of the diffusion dynamics with respect to the spine. The custom simulation code utilizes a diffusion equation for the dendrite based on a model by \cite{Biess2011}, and a time-dependent ordinary differential equation for the spine. Thereby, it does not consider spatial diffusion between spine and dendrite but instead features rate factors that govern calcium exchange between the two segments. In contrast to that, Arbor considers diffusion throughout the whole morphological structure, including the spines. The custom code, however, incorporates distinct influx and outflux coefficients inspired by \cite{Biess2011} and \cite{schuss2007narrow}. 
To maintain consistency, we neglected the possibility of different rates in the custom code and used a unified rate for the diffusion between the dendrite and the spines. Accordingly, we needed to adjust the amplitude of the injected current in the custom code to align with Arbor (cf. Table \ref{tab:parameters_model_heterosyn_calcium}).


\subsection{Synaptic tagging and capture, in individual synapses and in networks of single-compartment neurons}\label{ssec:stc_synapse_and_network}

The early and late phase, i.e., the induction and maintenance, of long-term synaptic plasticity are described by the so-called \emph{synaptic tagging and capture} (STC) hypothesis \citep{FreyMorris1997,RedondoMorris2011}.
As a next step for our modeling demonstrations, we reproduce the results of standard protocols eliciting early- and late-phase plasticity at a single nerve fiber or single synapse (cf. the experimental results in \cite{Sajikumar2005}). To achieve this, we implemented the complex theoretical model from \cite{LuboeinskiTetzlaff2021} in Arbor, requiring all of the new core components that we introduced in section \ref{sec:core}. We cross-validated our Arbor implementation by comparing its results to results from a stand-alone simulator for synaptic memory consolidation written in C\verb!++! \cite{StandAloneCode} that was custom-developed and used in the scope of several previous studies \citep{LuboeinskiTetzlaff2021,LuboeinskiTetzlaff2022,Luboeinski2021thesis,Lehr2022}.
Note that since the stand-alone simulator considers idealized point-neuron dynamics, we also implemented an approximate point neuron in Arbor, by integration of the current flow over the surface of a very small cylinder (cf. Table \ref{tab:parameters_model_stc}).

In the following, we provide the mathematical description of the used plasticity model. 
The parameter values can be found in Table~\ref{tab:parameters_model_stc}. For the other parts of the model, please refer to the code or the original studies \citep{LuboeinskiTetzlaff2021, Lehr2022}. Also note that a UML sequence diagram of the model implementation is provided in Supplementary Fig.~S5.

The total synaptic weight
\begin{equation}
w = h + h_0 \cdot z \label{eq:total_syn_weight}
\end{equation}
consists of two variable contributions, accounting for the two-phase nature of STC mechanisms. The first contribution is given by the early-phase weight $h$, while the second one is the late-phase weight $z$. The factor $h_0$ is used to normalize $z$ such that it has the same dimension as $h$.
The early-phase weight is described by the following differential equation:

\begin{align}
\tau_h \frac{dh(t)}{dt} = 0.1\,(h_0 - h(t)) + \gamma_\textrm{p}\cdot(10\,\textrm{mV}-h(t))\cdot\Theta\left[c(t)-\theta_\textrm{p}\right]\nonumber\\ 
- \gamma_\textrm{d}\cdot h(t)\cdot\Theta\left[c(t)-\theta_\textrm{d}\right] + \xi(t),\label{eq:early_phase_weight}
\end{align}
where $\tau_h$ is a time constant, $\gamma_\textrm{p}$ is the potentiation rate, $\gamma_\textrm{d}$ is the depression rate, and $c(t)$ is the calcium concentration at the postsynaptic site. Finally, $\xi(t)$ constitutes a noise term that depends on the occurrence of potentiation or depression:

\begin{align}
\xi(t) = \sqrt{\tau_h \left(\Theta\left[c(t)-\theta_\textrm{p}\right] + \Theta\left[c(t)-\theta_\textrm{d}\right]\right)}\cdot\sigma_\textrm{pl}\cdot\Gamma(t) \label{eq:plasticity_noise_2}
\end{align}
with scaling factor $\sigma_\textrm{pl}$ and Gaussian white noise $\Gamma(t)$ (which has a mean value of zero and a variance of $1/dt$ \citep{Gillespie1996}).
Note that this calcium-driven model of early-phase plasticity is based on the model by Graupner \& Brunel \cite{GraupnerBrunel2012}, which we considered in subsection \ref{ssec:ca_plast_graupner_brunel}.
Adaptations by \cite{Li2016} and \citep{LuboeinskiTetzlaff2021} 
have enabled the model to be compatible with synaptic tagging and capture models (also see \citep{Luboeinski2021thesis,LuboeinskiTetzlaff2024}).
As in the original model (cf. Eq.~\ref{eq:ca_model_graupner_brunel}), the calcium concentration depends on pre- and postsynaptic spikes at times $t^n_\textrm{pre}$ and $t^m_\textrm{post}$, and is described by the following equation:

\begin{equation}
\frac{dc(t)}{dt} = -\frac{c(t)}{\tau_\textrm{c}} + c_\textrm{pre}\cdot\sum\limits_{n}\delta(t-t^n_\textrm{pre}-t_\textrm{c,delay}) + c_\textrm{post}\cdot\sum\limits_{m}\delta(t-t^m_\textrm{post}),\label{eq:ca_for_stc_model}
\end{equation}
where $\tau_c$ is a time constant, $c_\textrm{pre}$ and $c_\textrm{post}$ are the spike-induced increases in the calcium concentration, $t_\textrm{c,delay}$ is the delay of the presynaptic contributions, and $\delta(\cdot)$ is the Dirac delta distribution. Note that calcium does not have a particular dimension here since it is only considered in the point-approximated synapses. The late-phase synaptic weight, which depends on the early-phase weight $h(t)$, is given by \citep{Li2016}:

\begin{align}
{\tau_{z}}\frac{dz(t)}{dt} = & \quad p(t)\cdot f_\textrm{int} \cdot (1-z(t))\cdot\Theta\left[(h(t)-h_0)-\theta_\textrm{tag}\right]\nonumber\\
& - p(t)\cdot f_\textrm{int} \cdot (z + 0.5)\cdot\Theta\left[(h_0-h(t))-\theta_\textrm{tag}\right],\label{eq:late_phase_weight}
\end{align}
where $\tau_z$ is a time constant, $p(t)$ is the concentration of plasticity-related products or proteins (PRPs), $f_\textrm{int}$ accounts for the integration of PRPs into the synaptic structure, and $\theta_\textrm{tag}$ is the tagging threshold. The synapse is considered tagged if the change in early-phase weight $\left|h(t)-h_0\right|$ exceeds the tagging threshold. Late-phase potentiation or depression occurs when the synapse is both tagged and PRPs are abundant ($p(t) > 0$).
The synthesis of PRPs depends on another threshold crossing and is described by \citep{Clopath2008,Li2016,LuboeinskiTetzlaff2021}:

\begin{equation}
\tau_p \frac{dp(t)}{dt} = -p(t) + p_\textrm{max}\cdot\Theta\left[\left(\sum\limits_\textrm{synapses}\left|h(t)-h_0\right|\right)-\theta_\textrm{pro}\right] \label{eq:prp_single_comp}
\end{equation}

with time constant $\tau_p$, the PRP synthesis threshold $\theta_\textrm{pro}$, and the PRP synthesis scaling constant $p_\textrm{max}$. Note that for a single synapse, the sum in the threshold condition reduces to the early-phase weight change of that individual synapse only.

As a first step for our model implementation, we considered basic dynamics of a single synapse with early-phase plasticity, and compared the results from Arbor and from the stand-alone simulator. The resulting curves match very well, as shown in Fig.~\ref{fig:stc_synapse_and_singlecomp_network}c,e,g (also see Supplementary Fig.~S2). 
To rule out any significant deviations that might be caused by the different numerical methods used by the two simulators, we further checked the validity of both approaches by comparing to a Brian 2 implementation (see Supplementary Fig.~S3; since Brian 2 comes with a Heun solver, which can solve stochastic differential equations that contain multiplicative noise with very high precision, it can provide a benchmark for the accuracy of other simulations).

Next, we considered basic dynamics of a single synapse with late-phase plasticity, which is shown in Fig.~\ref{fig:stc_synapse_and_singlecomp_network}d,f. Under continuous strong stimulation, the early-phase weight reaches its maximum after some time, and the late-phase weight subsequently converges to roughly the same value. Eventually, the early-phase weight decays. Again, we compared the results that we obtained from our Arbor implementation with the stand-alone simulator, finding the curves to match very well. In addition, we compared to Brian 2 again, which also shows a very good match (shown in Supplementary Fig.~S3).

Following the implementation of the basic dynamics of early- and late-phase synaptic plasticity, we used Arbor to reproduce the outcome of experimental standard protocols for early- and late-phase plasticity.
Again, our results obtained with Arbor are in agreement with the results from the stand-alone simulator (see Supplementary Fig.~S6). 
In summary, by matching the aforementioned results, we could prove the validity of our Arbor model implementation (as well as the validity of the stand-alone and Brian 2 implementations \citep{StandAloneCode,Brian2N1SCode}) with respect to the simulation of early- and late-phase synaptic plasticity.

\begin{figure}[!ht]
	\centering
    \includegraphics[width=\textwidth]{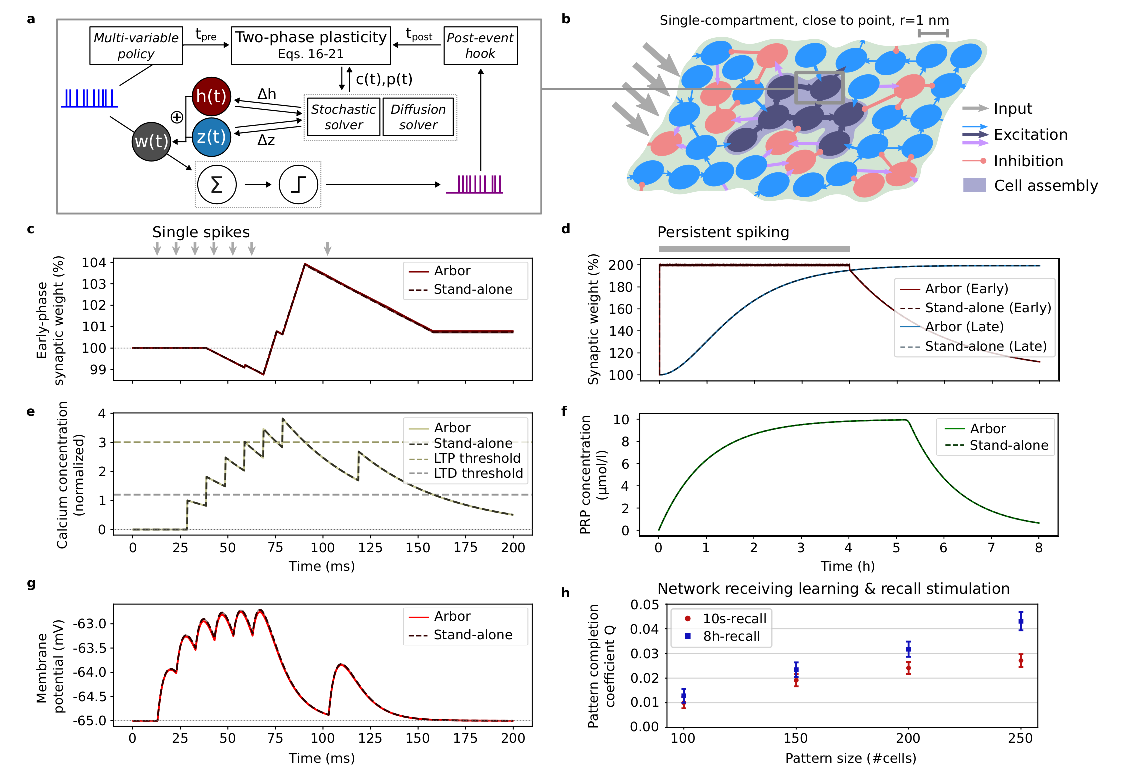}
    \caption{\small \textbf{Basic early- and late-phase plasticity with synaptic tagging and capture (STC), cross-validated with stand-alone simulator, and memory recall performance with single-compartment model.} 
    \textbf{(a)} Paradigm of two-phase synaptic plasticity with calcium-based early phase and late phase described by synaptic tagging and capture (see \citep{LuboeinskiTetzlaff2021}). Specific spiking input drives the weight dynamics, which further depend on stochastic dynamics and diffusion of PRP (results shown in (c--g)). New features of the Arbor core code are highlighted in italic. \textbf{(b)} Fraction of a neuronal network consisting of excitatory (blue and dark blue circles) and inhibitory neurons (red circles). Following external input, the synapses between excitatory neurons undergo plastic changes implemented as detailed in (a), forming a Hebbian cell assembly (related results in (h)). \textbf{(c)} Averaged noisy early-phase synaptic weight (cf. Eq.~\ref{eq:early_phase_weight}). The synapse receives spiking input at pre-defined times (indicated by bold gray arrows). Goodness of fit between the mean curves: $\mathrm{CV} = 0.999$, $\mathrm{RMSE} = 0.040\unit{mV}$. 
    \textbf{(d)} Limit cases of early- and late-phase synaptic weight (cf. Eqs.~\ref{eq:early_phase_weight} and \ref{eq:late_phase_weight}). The presynaptic neuron is stimulated to spike at maximal rate (indicated by gray bar). The late-phase weight has been shifted for graphical reasons (cf. Eq.~\ref{eq:late_phase_weight}; early phase: $\mathrm{CV} = 0.201$, $\mathrm{RMSE} = 0.226\unit{mV}$; late phase: $\mathrm{CV} > 0.999$, $\mathrm{RMSE} = 0.055\unit{mV}$). 
    \textbf{(e)} Postsynaptic calcium concentration, successively crossing the thresholds for depression (LTD) and potentiation (LTP) (cf. Eq.~\ref{eq:ca_for_stc_model}; $\mathrm{CV} > 0.999$, $\mathrm{RMSE} = 0.065$). 
    \textbf{(f)} The postsynaptic PRP concentration rises until it reaches its maximum through the continued stimulation (cf. Eq.~\ref{eq:prp_single_comp}; $\mathrm{CV} = 0.998$, $\mathrm{RMSE} = 0.107\unit{{\textmu}mol/l}$). 
    \textbf{(g)} Membrane potential of the postsynaptic neuron ($\mathrm{CV} > 0.999$, $\mathrm{RMSE} = 0.151\unit{mV}$). Basic early-phase plasticity dynamics (c,e,g): average across $10$ batches, each consisting of $100$ trials. Baseline levels are represented by fine, dotted lines. Basic late-phase plasticity dynamics (d,f): average across $10$ batches, each consisting of $10$ trials. Noise seeds were drawn independently for each trial. Results of Arbor are represented by continuous lines, results of the stand-alone simulator \cite{StandAloneCode} by darker, dashed lines. For each curve, error bands represent the standard error of the mean (mostly too small to be visible).
    \textbf{(f)} Memory recall in networks of single-compartment neurons simulated with Arbor (qualitatively reproducing the point-neuron results of \cite{LuboeinskiTetzlaff2021}). Pattern completion is measured by the coefficient $Q$ (see Eq.~\ref{eq:recall_Q}) for stimulated patterns of varied size (a varied number of neurons are stimulated for learning/recall). Average over $100$ network realizations; error bars represent the 95\% confidence interval.
    \label{fig:stc_synapse_and_singlecomp_network}}
\end{figure}

Next, as a major test for the newly implemented plasticity functionality in Arbor, we employed a recurrent spiking neural network model comprising plastic synapses (which had before been used in \cite{LuboeinskiTetzlaff2021}). The network consists of $2000$ single-compartment neurons that exhibit synaptic connections with a probability of $0.1$. A fraction of $1600$ of these neurons are excitatory. The synapses within this excitatory population are plastic and follow the model that we considered above (Eqs.~\ref{eq:total_syn_weight}--\ref{eq:prp_single_comp}). The remaining neurons are inhibitory and connected via static synapses. See Supplementary Fig.~S4 for cross-validation plots of the resulting spike transmission.

Using this network model, we aimed to simulate the `10s-recall' and `8h-recall' paradigms that had been investigated in \cite{LuboeinskiTetzlaff2021}. These paradigms comprise the formation of a cell assembly that retains the memory of a given learning stimulus, and the recall of this memory upon stimulation of half of the original neurons either $10\unit{s}$ or $8\unit{h}$ later. Here, we chose groups of $100$, $150$, $200$, or $250$ neurons receiving the learning stimulus to form the `core' of the learned cell assembly. Subsequently, this cell assembly could become consolidated by the synaptic tagging and capture mechanisms, which we tested via recall stimulation after $8\unit{h}$. Simulating such long biological time spans in reasonable compute time required us to implement a `fast-forward' computation mechanism for phases of slow network dynamics. For this, we first implemented a state-saving mechanism, which enables to stop the Arbor simulation at an arbitrary point, save the synaptic weights and PRP concentrations, and then set up a new Arbor \emph{recipe} with the previous state (note that Arbor already provides a checkpointing feature, but this does not go as far as to enable changing parts of the recipe). The simulation is then continued with long timesteps for the slow network dynamics without considering spiking and calcium dynamics, which results in much shorter compute time (also cf. runtime results in subsection \ref{ssec:benchmarking} and Supplementary Fig.~S13). Finally, before performing memory recall, the compute mode is switched again to simulate the network dynamics in full detail. Note that a similar approach had previously been taken with the stand-alone simulator \cite{LuboeinskiTetzlaff2021,LuboeinskiTetzlaff2024}.

For the plasticity dynamics, we followed the formulation presented above (Eqs.~\ref{eq:total_syn_weight}--\ref{eq:prp_single_comp}), but since the considered neurons did now hold multiple synapses, an adaptation of the technical implementation became necessary to simulate global PRP dynamics. Specifically, we could no longer compute the PRP dynamics in the NMODL mechanisms of the synapses (cf. the code in \citep{Arbor2N1SCode} and \citep{ArborNetworkScCode}). Instead, we had to compute the weight change sum from Eq.~\ref{eq:prp_single_comp} in the soma. For this, we needed to model the signaling of the weight changes from the synapses to the soma. We did this by implementing a putative substance that diffuses across the compartments of the neuron. We call this substance the `signal triggering PRP synthesis' (SPS). In the case of a single-compartment neuron, naturally, the SPS reaches the soma instantaneously (the multi-compartment case is considered in subsection \ref{ssec:stc_morpho_neuron_network}). 
In every timestep, the amount of SPS is compared against the threshold $\theta_\textrm{pro}$ (cf. Eq.~\ref{eq:prp_single_comp}), letting PRP synthesis take place as long as the threshold is crossed. 
Subsequently, produced PRPs diffuse across the neuron to reach the synapses, where they can give rise to late-phase weight changes. 

Finally, to measure the performance in recalling the input pattern defined by the learning stimulus, we use the following quantity \citep{LuboeinskiTetzlaff2021}:

\begin{equation}
Q := \frac{\bar{\nu}_\text{ans} - \bar{\nu}_\text{ctrl}}{\bar{\nu}_\text{as}}.\label{eq:recall_Q}
\end{equation}

For this, the population of excitatory neurons is divided into three subpopulations: assembly neurons that are stimulated by both recall and learning stimulus (`as'), assembly neurons that are not stimulated by recall but were stimulated by learning stimulus (`ans'), and control neurons that are stimulated by neither recall nor learning stimulus (`ctrl'). The mean firing rates in the three subpopulations upon 10s- and 8h-recall, computed using time windows of $0.5\,\text{s}$ centered at $t_\text{recall}=20.1\,\text{s}$ and $t_\text{recall}=28810.1\,\text{s}$, are denoted by $\bar{\nu}_{\text{as}}$, $\bar{\nu}_{\text{ans}}$, and $\bar{\nu}_{\text{ctrl}}$, respectively. Thus, values of $Q > 0$ indicate that the pattern is successfully recalled.

The qualitative reproduction of the results from \cite{LuboeinskiTetzlaff2021} with our Arbor implementation is shown in Figure \ref{fig:stc_synapse_and_singlecomp_network}h (also cf. Supplementary Fig.~S10).
While we previously found that elementary dynamics of the used plasticity rule match very well for Arbor, the stand-alone simulator, and Brian 2 (Fig.~\ref{fig:stc_synapse_and_singlecomp_network}c--g and Supplementary Figs. S2--S4), the behavior of the full network can not be reproduced in full detail. We attribute this to three factors: 
First, the different simulators use different numerical solving methods. Second, the neurons in the stand-alone simulator and in Brian 2 are actual point neurons, whereas in Arbor we only consider \emph{approximate} point neurons, which are described by a very small but finite-sized cylinder. 
And finally, the high complexity of the network dynamics further amplifies the existing differences between the simulators. 
Thus, although the qualitative behavior is maintained, not all quantitative deviations can be eliminated.

\begin{table}[!ht]
\small
\begin{center}
\begin{tabular}{|l|p{2.5cm}|p{6.0cm}|p{2.5cm}|}
    \hline {Symbol} & {Value} & {Description} & {Refs.} \\ 
	\hline
	\hline {$h_0$} & {$4.20075\unit{mV}$ $= 0.5\,\frac{\gamma_\textrm{p}}{\gamma_\textrm{p} + \gamma_\textrm{d}}$} & {Baseline value of the excitatory$\rightarrow$excitatory coupling strength} & {\cite{GraupnerBrunel2012,Li2016,LuboeinskiTetzlaff2021}}\\
	\hline{$t_\textrm{c,delay}$} & {$0.0188\unit{s}$} & {Delay of postsynaptic calcium influx after presynaptic spike} & {\cite{GraupnerBrunel2012,Li2016,LuboeinskiTetzlaff2021}}\\
	\hline {$c_\textrm{pre}$} & {$1.0$ ($0.6$)} & {Presynaptic calcium contribution (in vivo adjusted)} & {\cite{GraupnerBrunel2012,Li2016,Higgins2014,LuboeinskiTetzlaff2021}}\\
	\hline {$c_\textrm{post}$} & {$0.2758$ ($0.1655$)} & {Postsynaptic calcium contribution (in vivo adjusted)} & {\cite{GraupnerBrunel2012,Li2016,Higgins2014,LuboeinskiTetzlaff2021}}\\
	\hline {$\tau_c$} & {$0.0488\unit{s}$} & {Calcium time constant} & {\cite{GraupnerBrunel2012,Li2016,LuboeinskiTetzlaff2021}}\\
	\hline {$\tau_h$} & {$688.4\unit{s}$} & {Early-phase time constant} & {\cite{GraupnerBrunel2012,Li2016,LuboeinskiTetzlaff2021}}\\
	\hline {$\tau_p$} & {$60\unit{min}$} & {PRP time constant} & {\cite{Clopath2008,Li2016,LuboeinskiTetzlaff2021}}\\
	\hline {$\tau_{z}$} & {$60\unit{min}$} & {Late-phase time constant} & {\cite{Clopath2008,Li2016,LuboeinskiTetzlaff2021}}\\
	\hline {$\gamma_\textrm{p}$} & {$1645.6$} & {Potentiation rate} & {\cite{GraupnerBrunel2012,Li2016,LuboeinskiTetzlaff2021}}\\
	\hline {$\gamma_\textrm{d}$} & {$313.1$} & {Depression rate} & {\cite{GraupnerBrunel2012,Li2016,LuboeinskiTetzlaff2021}}\\
	\hline {$\theta_\textrm{p}$} & {$3.0$} & {Calcium threshold for potentiation} & {\cite{Li2016,LuboeinskiTetzlaff2021}}\\
	\hline {$\theta_\textrm{d}$} & {$1.2$} & {Calcium threshold for depression} & {\cite{Li2016,LuboeinskiTetzlaff2021}}\\
	\hline {$\sigma_\textrm{pl}$} & {$2.90436\unit{mV}$} & {Standard deviation for plasticity fluctuations} & {\cite{GraupnerBrunel2012,Li2016,LuboeinskiTetzlaff2021}}\\
	\hline {$p_\textrm{max}$} & {$10.0\unit{{\textmu}mol/l}$} & {PRP synthesis scaling constant (equilibrium PRP concentration under ongoing PRP synthesis)} & {\cite{Clopath2008,Li2016,LuboeinskiTetzlaff2021}}\\
	\hline {$\theta_\textrm{pro}$} & {$2.10037\unit{mV}$ $= 0.5\,h_0$} & {PRP synthesis threshold} & {\cite{Li2016,LuboeinskiTetzlaff2021}}\\
	\hline {$\theta_\textrm{tag}$} & {$0.840149\unit{mV}$ $= 0.2\,h_0$} & {Tagging threshold} & {\cite{Li2016,LuboeinskiTetzlaff2021}}\\
	\hline {$f_\textrm{int}$} & {$0.1\unit{l/{\textmu}mol}$} & {Late-phase factor accounting for PRP integration into the synapse} & {This study}\\
	\hline {$r_\textrm{comp}$} & {$1{\cdot}10^{-3}\unit{{\textmu}m}$} & {Radius of the single-compartment cell} & {This study}\\
	\hline {$l_\textrm{cell}$} & {$2{\cdot}10^{-3}\unit{{\textmu}m}$} & {Length of the single-compartment cell} & {This study}\\
	\hline
\end{tabular}
\caption{\small Parameters for the model with calcium-based early-phase plasticity and STC-based late-phase plasticity based on \citep{Li2016} and \citep{LuboeinskiTetzlaff2021}. The calcium concentration in this model is a dimensionless quantity since it is only considered in the synapses (see main text). We use parameters for the calcium-based early-phase model that were fitted on hippocampal slice data \citep{WittenbergWang2006,GraupnerBrunel2012}. For networks, the calcium contribution parameters are corrected by a factor of $0.6$ to account for in vivo conditions (cf.~\citep{Higgins2014}).\label{tab:parameters_model_stc}}
\end{center}
\end{table}


\subsection{Synaptic memory consolidation in networks of morphological neurons}\label{ssec:stc_morpho_neuron_network}

Now, we are finally going to demonstrate how we can exploit Arbor's capabilities to simulate networks of multi-compartment neurons with synaptic plasticity. To this end, we extend the size of the cylindrical compartment considered in the previous subsection, split it into two cylinders, add to their middle a compartment for PRP synthesis, and use this as the soma (see Fig.~\ref{fig:stc_multicomp_network}a,b).
We further add two cylinders to represent dendritic branches -- one to account for an apical dendrite and one to account for basal dendrites. These branches will have synapses at their tips and are meant to approximate the impact of apical and basal dendritic input onto the soma. The parameters of the morphology are given in Table \ref{tab:parameters_morpho_neuron}. 
For simplification, all compartments have the same diameter, and we chose the diameter value to yield biologically realistic functional dynamics (cf. \citep{Jiang2020,CarnevaleHines2006}).

The network is structured such that the apical dendrites receive external input, while the basal dendrites account for the recurrent connectivity of the excitatory neurons within the simulated network. This is grounded by findings on the neocortical layer structure of the neocortex, where basal dendrites receive the inputs from within a layer and apical dendrites receive inputs from other layers \cite{Larkman1991,Gillon2023}. 
Finally, the inhibitory neurons form connections directly onto the soma. To account for the propagation of excitatory postsynaptic potentials along the morphology of the basal dendrites, we introduced a correction factor $c_\textrm{morpho}$. Note that otherwise, the electrical properties of the neurons are the same as in the single-compartment case presented in the previous subsection \ref{ssec:stc_synapse_and_network}. 

To implement the plasticity dynamics, as in the previous subsection, we use again a putative SPS substance that diffuses from the synapses across the whole neuron with the purpose to signal weight changes to the soma. 
For the sake of simplicity, we assume that the diffusion of the SPS towards the soma happens very fast, with a diffusivity of $D_\textrm{sps}=10^{-11}\unit{m\textsuperscript{2}/s}$. 
In every timestep, we compare the concentration of the SPS in the center of the soma against a PRP synthesis threshold (cf.~Eq.~\ref{eq:prp_single_comp}). PRP synthesis will take place as long as the threshold is crossed. 
Howsoever, note that here we compare to the \emph{concentration} and not to the \emph{amount} of SPS. This enables a more efficient NMODL implementation, but 
requires the renormalization of the threshold parameter by scaling it with the total volume of the neuron:

\begin{equation}
\theta_\textrm{pro}^{*} = \frac{\theta_\textrm{pro}}{V_\textrm{tot}} = \frac{\theta_\textrm{pro}}{\sum_i V_i}, \label{eq:theta_pro_multi_comp}
\end{equation}
where $V_i$ are the volumes of the individual compartments (the neurons used here comprise up to $48$ compartments). 
Also note that unless the diffusion happens instantaneously, in the multi-compartment case, the SPS concentration measured in the soma will never perfectly reflect the total amount of SPS in the whole neuron, which constitutes an essential difference to the single-compartment case. 

For the diffusion of PRPs within the simulated neurons, we again use the same mechanism as described in the previous subsection \ref{ssec:stc_synapse_and_network}. However, here we simulate `real' diffusion across a spatial morphology structure (cf. Eq.~\ref{eq:diffusion}), for which we decided to consider three diffusivity values (see Fig.~\ref{fig:stc_multicomp_network}f--h,j--l). For the fastest considered  diffusion ($D_\textrm{p}=10^{-11}\unit{m\textsuperscript{2}/s}$), the PRPs reach all parts of the neuron almost instantaneously, such that there is no difference to a single-compartment model in this respect (cf. the distribution of PRPs in Supplementary Fig.~S8). However, although we adjusted the $c_\text{morpho}$ parameter for the postsynaptic potentials to match those in the single-compartment paradigm, the electrical properties of the morphological structure show to have an impact on the firing rate of the neurons. This is demonstrated by the results for single- and multi-compartment neurons on memory recall after $10$ seconds, which naturally does not depend on PRPs (Figs.~\ref{fig:stc_synapse_and_singlecomp_network}h and \ref{fig:stc_multicomp_network}e,i).  
On the other hand, considering the long-term dynamics with very slow diffusion ($D_\textrm{p}=10^{-19}\unit{m\textsuperscript{2}/s}$), the PRPs will not reach the target synapses within the time window of the synaptic tag and thus cannot elicit late-phase plasticity. For moderate diffusion values ($D_\textrm{p}=10^{-15}\unit{m\textsuperscript{2}/s}$), PRPs reach the synapses after a certain time that arises from a complex interplay between synapses, soma, and dendrites (see the spatial distribution in Supplementary Fig.~S9). These dynamics still enable functional memory recall and can serve to regulate the late-phase maintenance of synaptic changes. 
For increased size of the dendrites, however, the memory recall performance tends to become worse (see Fig.~\ref{fig:stc_multicomp_network}f--h,j--l; also cf. the results for the mutual information in Supplementary Fig.~S11). 
Interestingly, increasing the size of the neurons (measured by the diameter of the soma and dendrites) has a converse effect: the memory recall is improved (Fig.~\ref{fig:stc_multicomp_network}j--l).

Although our present study has been focused on simulation methods, these last findings may provide interesting theoretical insights into the role of neuronal structure and dynamics for cognitive functionality at the network level. As the presented results show, Arbor enables to seamlessly move from single- to multi-compartment neurons in a complex network model, leaving the remaining parts of the model unchanged. In the future, the framework that we have developed can be used as a basis for further investigations on neural networks involving diffusion dynamics in multi-compartment neurons.

\begin{table}[!ht]
\small
\begin{center}
\begin{tabular}{|p{1.5cm}|p{1.5cm}|p{1.0cm}|p{6.0cm}|p{2.5cm}|}
    \hline {Paradigm} & {Symbol} & {Value} & {Description} & {Refs.} \\
	\hline
	\hline
	{(Any)} & {$\Delta{l}_\textrm{comp}$} & {$1.0\unit{\textmu{m}}$} & {Length of one compartment} & {This study}\\
	\hline
	\multirow{3}*{\parbox{1.5cm}{Small cells}} & {$r_\textrm{comp}$} & {$6.0\unit{\textmu{m}}$} & {Effective radius of a compartment (used for dendrites as well as soma)} & {\citep{Jiang2020,vanAerde2015,Dzaja2014,CarnevaleHines2006}}\\
	\cline{2-5}
	{} & {$l_\textrm{soma}$} & {$12.0\unit{\textmu{m}}$} & {Length of the soma} & {\citep{Jiang2020,vanAerde2015,Dzaja2014}}\\
	\hline
	\multirow{3}*{\parbox{1.5cm}{Large cells}} & {$r_\textrm{comp}$} & {$12.0\unit{\textmu{m}}$} & {Effective radius of a compartment (used for dendrites as well as soma)} & {\citep{Jiang2020,vanAerde2015,Dzaja2014,CarnevaleHines2006}}\\
	\cline{2-5}
	{} & {$l_\textrm{soma}$} & {$24.0\unit{\textmu{m}}$} & {Length of the soma} & {\citep{Jiang2020,vanAerde2015,Dzaja2014}}\\
	\hline
	\multirow{3}*{\parbox{1.5cm}{Small dendrites}} & {$l_\textrm{dendriteA}$} & {$12.5\unit{\textmu{m}}$} & {Length of apical dendritic branch} & {\citep{CarribaDavies2017,vanAerde2015,CarnevaleHines2006}}\\
	\cline{2-5}
	{} & {$l_\textrm{dendriteB}$} & {$5.0\unit{\textmu{m}}$} & {Length of basal dendritic branch} & {\citep{CarribaDavies2017,vanAerde2015,CarnevaleHines2006}}\\
	\cline{2-5}
	{} & {$c_\textrm{morpho}$} & {$1.035$} & {Correction factor for the altered impact of postsynaptic potentials due to the morphology} & {This study (referring to model in \citep{LuboeinskiTetzlaff2021})}\\
	\hline
	\multirow{3}*{\parbox{1.5cm}{Large dendrites}} & {$l_\textrm{dendriteA}$} & {$25.0\unit{\textmu{m}}$} & {Length of apical dendritic branch} & {\citep{CarribaDavies2017,vanAerde2015,CarnevaleHines2006}}\\
	\cline{2-5}
	{} & {$l_\textrm{dendriteB}$} & {$10.0\unit{\textmu{m}}$} & {Length of basal dendritic branch} & {\citep{CarribaDavies2017,vanAerde2015,CarnevaleHines2006}}\\
	\cline{2-5}
	{} & {$c_\textrm{morpho}$} & {$1.020$} & {Correction factor for the altered impact of postsynaptic potentials due to the morphology} & {This study (referring to model in \citep{LuboeinskiTetzlaff2021})}\\
	\hline
\end{tabular}
\caption{\small Cell morphology parameters for the network simulations of memory formation and consolidation with morphological neurons (subsection \ref{ssec:stc_morpho_neuron_network}). We investigated each combination of the cell and dendrite sizes. The values are chosen to approximate the effective functional dynamics that arise from the structures of real neurons (essentially, pyramidal cells) in hippocampus or neocortex. See the main text for more details.\label{tab:parameters_morpho_neuron}}
\end{center}
\end{table}

\clearpage

\begin{figure}[!ht]
    \captionsetup{labelformat=original}
	\centering
    \includegraphics[width=\textwidth]{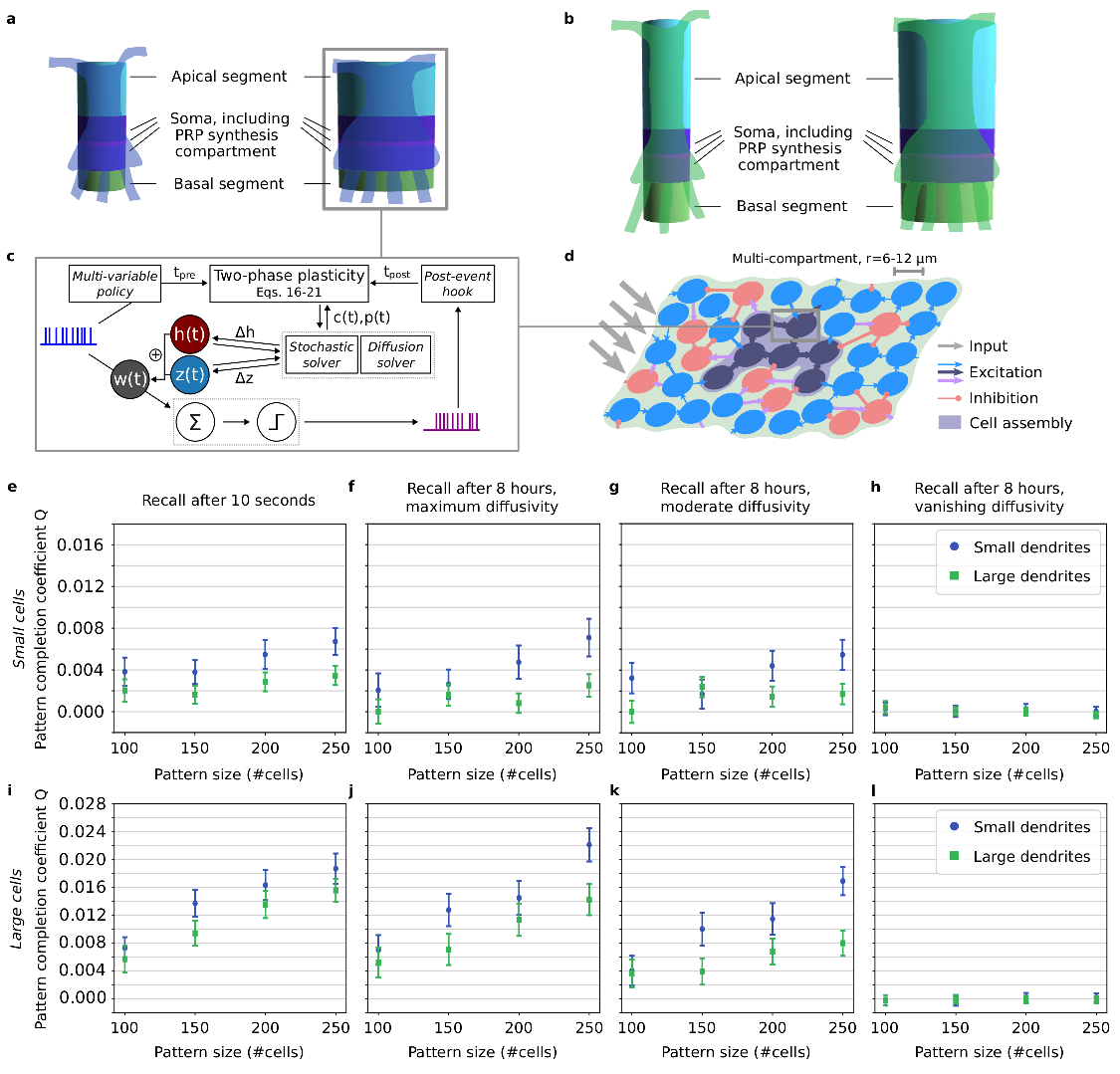}
    \caption{\small \textbf{Memory recall in a recurrent network of multi-compartment neurons after learning and after consolidation.} Results obtained with Arbor for networks of different kinds of multi-compartment neurons, demonstrating the impact of different values of the PRP diffusivity $D_\textrm{p}$ on memory consolidation. Networks consist of `small' cells (diameter of $6\unit{{\textmu}m}$) or of `large' cells (diameter of $12\unit{{\textmu}m}$), with either small or large dendrites (in which cases each neuron comprises in total $31$ or $48$ compartments, respectively). The diameter and length values are given in Table~\ref{tab:parameters_morpho_neuron}. \textbf{(a,b)} Illustrations of used cell structures, generated using \texttt{Arbor~GUI} \citep{Arbor-GUI-0-8-0}. Each segment is represented by a different color. A segment can consist of a multitude of compartments. Overlaid with illustrations of more realistic neuron structures that would have roughly similar functional properties. 
    \textbf{(a)} a small (left) and a large (right) cell with small dendrites, \textbf{(b)} the same with large dendrites (cf.~Table~\ref{tab:parameters_morpho_neuron}). \textbf{(c)} Paradigm of two-phase synaptic plasticity with calcium-based early phase and late phase described by synaptic tagging and capture. The impact of the diffusion of PRPs can be examined using different morphological neuron structures. New features of the Arbor core code are highlighted in italic. \textbf{(d)} Fraction of a neuronal network consisting of excitatory multi-compartment (blue and dark blue circles) and inhibitory neurons (red circles). Following external input, the synapses between excitatory neurons undergo plastic changes implemented as detailed in (c), forming a Hebbian cell assembly (related results in (e--l)).\label{fig:stc_multicomp_network}}
\end{figure}
\clearpage
\begin{figure}
    \captionsetup{labelformat=adja-page}
    \ContinuedFloat
    \caption{\small \textbf{(e-h)} Memory recall measured by pattern completion coefficient $Q$ (see Eq.~\ref{eq:recall_Q}) for a stimulated subset of varied size (a varied pattern of neurons are stimulated for learning/recall). Value $Q>0$ indicates the successful recall of a memory representation. Average over $100$ network realizations. Error bars represent the 95\% confidence interval. \textbf{(e)} Recall stimulation at $10\unit{s}$ after learning (technically, $D_\textrm{p}=10^{-11}\unit{m\textsuperscript{2}/s}$, but late-phase plasticity does not occur on such short timescales). \textbf{(f)} Recall stimulation at $8\unit{h}$ after learning, $D_\textrm{p}=10^{-11}\unit{m\textsuperscript{2}/s}$. \textbf{(g)} Recall stimulation at $8\unit{h}$ after learning, $D_\textrm{p}=10^{-15}\unit{m\textsuperscript{2}/s}$. \textbf{(h)} Recall stimulation at $8\unit{h}$ after learning, $D_\textrm{p}=10^{-19}\unit{m\textsuperscript{2}/s}$. \textbf{(i-l)} Same as (e-h), but for large cells that consist of segments of twice the diameter.}
\end{figure}


\subsection{Runtime and memory benchmarking with the synaptic memory consolidation model}\label{ssec:benchmarking}

Employing the network model introduced in the previous section, we scrutinize the resources required by Arbor to simulate networks of single- and multi-compartment neurons in different computing environments. To this end, we consider the networks of $2000$ single- or multi-compartment neurons with the 10s-recall paradigm that we used before (subsections \ref{ssec:stc_synapse_and_network} \& \ref{ssec:stc_morpho_neuron_network}). After running these network models with different Arbor backends on different hardware systems, we compare their runtime and memory use. In the single-compartment case, we also compare to Brian 2 with \texttt{cpp\_standalone} device \citep{BrianNetworkCode,Stimberg2019} as well as to the custom stand-alone simulator \citep{StandAloneCode} considered before.

\begin{figure}[!ht]
	\centering
    \includegraphics[width=\textwidth]{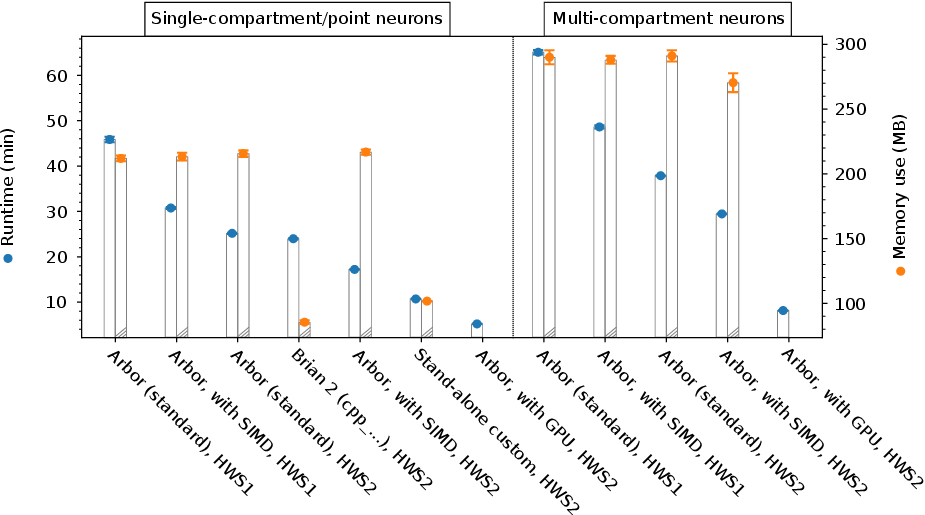}
    \caption{\small \textbf{Benchmarking results of runtime and memory use with the synaptic memory consolidation model in Arbor and point-neuron simulators.} For 10s-recall paradigm in networks of $2000$ neurons. The single-compartment simulations in Arbor as well the point-neuron simulations in Brian 2 (with \texttt{cpp\_standalone} device) \citep{BrianNetworkCode,Stimberg2019} and in the custom stand-alone simulator \citep{StandAloneCode} are conducted as described in subsection \ref{ssec:stc_synapse_and_network}; they are represented by data points on the left-hand side. The Arbor simulations with multi-compartment/morphological neurons of $48$ compartments are conducted as described in subsection \ref{ssec:stc_morpho_neuron_network} and represented by data points on the right-hand side. 
    Results are given for different hardware systems, HWS1: an older desktop computer (Intel Core i5-6600 CPU @ 3.30GHz, $1\times8\unit{GB}$ DDR3-RAM, using $1$ thread), HWS2: a newer compute server (AMD Ryzen Threadripper PRO 5995WX CPU, $8\times32\unit{GB}$ DDR4-RAM, using $1$ thread, in specified cases with NVIDIA T1000 $8\unit{GB}$ GPU). For Arbor, results are distinguished between standard CPU execution, CPU with SIMD support, and with GPU support. The respective left bars with blue data points show the total runtime of the simulations (comprising initialization and state propagation phases). Measurements were performed using \texttt{hyperfine} in version 1.15. The respective right bars with orange data points show the use of main memory, given by the maximum over time of the number of `dirty' bytes, including private and shared memory, as returned from \texttt{pmap}. Note that for the GPU cases considering the main memory use is not meaningful, since the GPU has its own memory.
    Data points represent the average over $10$ trials; error bars represent the standard deviation. Also cf. Supplementary Fig.~S12.\label{fig:benchmarking_memory_consolidation}}
\end{figure}

Fig.~\ref{fig:benchmarking_memory_consolidation} shows that the point-neuron simulators (Brian 2 and stand-alone) need shorter runtimes and less memory compared to Arbor. This is to be expected: Since point-neuron simulators consider neurons without any geometrical structure, they theoretically require much less calculation steps than Arbor, which accounts for at least one finite-size compartment per neuron. Furthermore, the custom stand-alone simulator is highly optimized for the particular model, and can therefore be thought to set an upper bound. Nevertheless, we found that Arbor's capability of employing GPU hardware can boost its runtime to be even faster than Brian 2 and the custom simulator (Fig.~\ref{fig:benchmarking_memory_consolidation}).
Another point to be mentioned is Arbor's support for single instruction, multiple data (SIMD) vectorization. As shown in Fig.~\ref{fig:benchmarking_memory_consolidation}, switching on SIMD vectorization provides a small but solid improvement in runtime. This comes without any cost for the end user, given that they have a CPU that supports SIMD, which has been the industry standard for many years. The only drawback of SIMD usage in Arbor might be its negative impact on code readability when developing custom mechanism code in C\verb!++!, which is rather a niche case.

Importantly, our results show that Arbor allows to shift from single-compartment neurons to morphological neurons at \emph{almost no cost}: both the runtime and the memory consumption only increase slightly when shifting from single-compartment neurons (Fig.~\ref{fig:benchmarking_memory_consolidation}, left-hand side) to neurons with $48$ compartments (Fig.~\ref{fig:benchmarking_memory_consolidation}, right-hand side). However, it is important to note that the considered multi-compartment network in general exhibits fewer spikes than the single-compartment version (Supplementary Fig.~S12b; also cf. subsection \ref{ssec:stc_morpho_neuron_network}). Thus, given that the number of spikes is a critical factor for the runtime, we checked if the runtime per spike follows as similar trend, and indeed found that this measure also exhibits a slight increase only (Supplementary Fig.~S12a).

Hence, while Arbor performs well in single-compartment neuron simulations, it excels in multi-compartment neuron models, providing all the necessary functionality to simulate morphological neurons with electrical cable properties and diffusing particles.

Finally, note that we also benchmarked results for 8h-recall simulations in Arbor (Supplementary Fig.~S13). The runtimes of those simulations are much longer than for 10s-recall (cf. Fig.~\ref{fig:benchmarking_memory_consolidation}). As we used the `fast-forward' approximation (see subsection \ref{ssec:stc_synapse_and_network}), there are only $10-20\%$ more timesteps than in the 10s-recall paradigm, so only a small fraction of the increase in runtime can be attributed to additional timesteps. Instead, we found the longer runtimes to be essentially due to a large overhead needed for switching between full and fast-forward computation, specifically, the setting of a large number of probes to store the state of the whole network. To counter this, in the future, we plan to augment our framework by introducing new mechanisms that serve to retain the simulation state.

\subsection{Runtime and memory benchmarking with large-scale networks}\label{ssec:benchmarking_very_large}

To underline our claims of Arbor's scalability and efficiency, in this final section, we consider a set of technical benchmarks based on the so-called busyring benchmark. This is a tunable, well-understood workload that enables stressing various parts of the underlying hardware architecture.
It has been used in this capacity as an acceptance benchmark for the novel JUPITER system at the Jülich Supercomputing Centre (JSC) \citep{Herten2024}.

\begin{figure}[t]
	\centering
    \includegraphics[width=\textwidth]{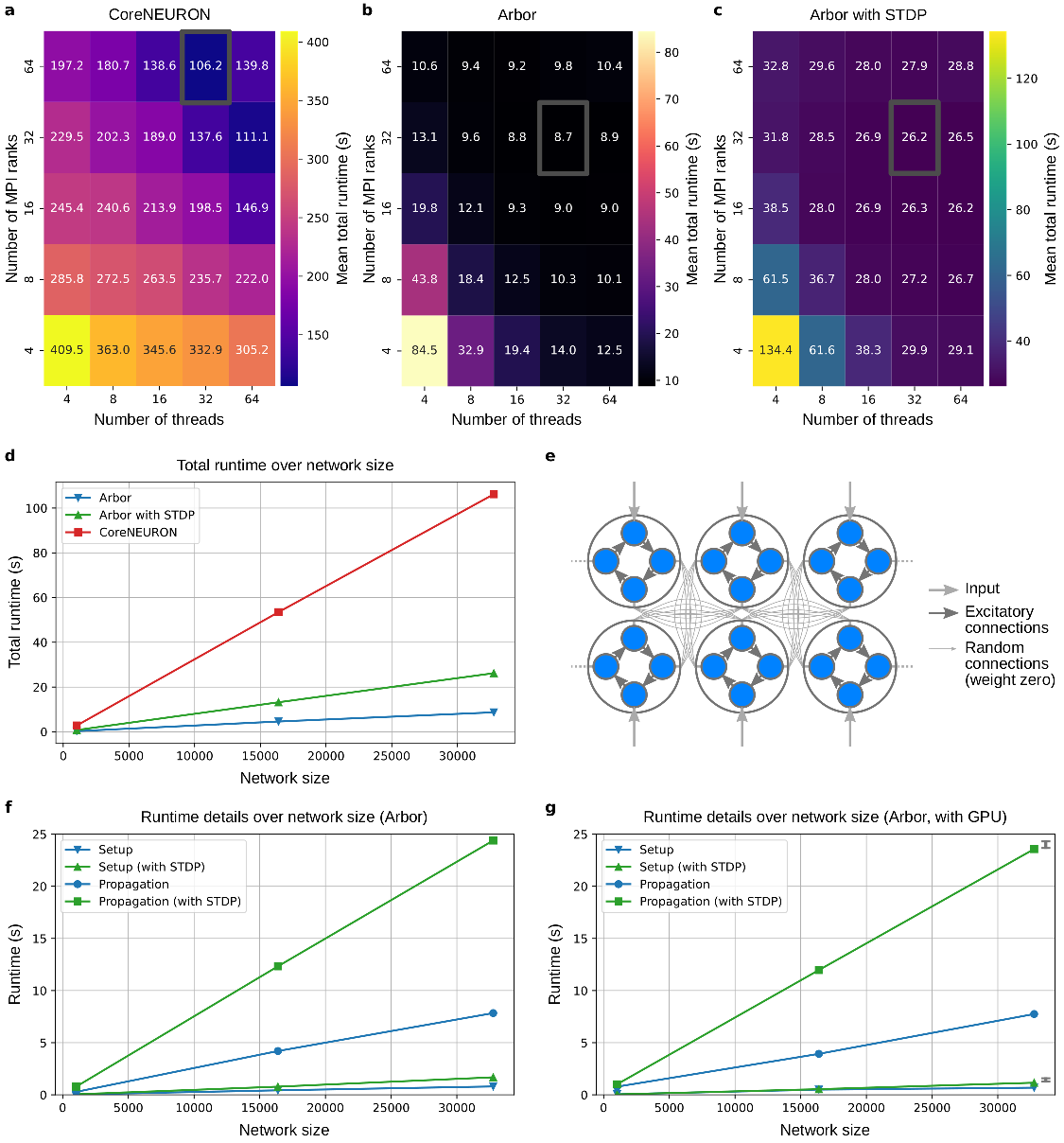}
    \caption{{\small \textbf{Benchmarking of simulation runtime for large-scale networks.} %
    \textbf{(a-c)} Total wallclock time to initialize and execute a simulation with $32768$ cells over $200\unit{ms}$ in Arbor and CoreNEURON. A busyring network of \texttt{simple branchy} cells with tree depth $2$ is used, run on the HWS2 system (AMD Ryzen Threadripper PRO 5995WX CPU with 64 cores, $8\times32\unit{GB}$ DDR4-RAM) with \textbf{(a)} CoreNEURON, \textbf{(b)} Arbor with SIMD, \textbf{(c)} Arbor with SIMD with STDP mechanisms for the random synapses. The respectively fastest paradigm for each implementation is highlighted by the gray box. \textbf{(d)} Scaling of the fastest results for the total runtime over network size. \textbf{(e)} Sketch of the busyring network consisting of rings of integrate-and-fire neurons (shown as blue disks), connected internally via excitatory synapses, and across the whole network via random synapses of weight zero. One neuron of each ring receives external stimulation. \textbf{(f)} Scaling of the setup and propagation runtime related to the total runtimes in \textbf{d}. \textbf{(g)} Scaling of the setup and propagation runtime for optimized total runtime using an additional NVIDIA T1000 8GB GPU. For the case with STDP, the GPU-mediated speedup is indicated by dark gray bars. All values are averaged over $10$ trials, with coefficient of variation $\mathrm{CV} < 0.06$ in all cases. See Tables \ref{tab:runtime_very_large} and \ref{tab:runtime_very_large_gpu} for more details.}\label{fig:benchmarking_runtime_very_large}}
\end{figure}

The basic unit of work in the busyring benchmark is a ring of $k$ neurons tuned to propagate a single spike indefinitely (see Fig.~\ref{fig:benchmarking_runtime_very_large}e).
Connections within the ring are realized via conductance-based exponential synapses with a uniform delay $t_\text{delay}$ and weight $w\neq 0$.
Thus, each neuron spikes at frequency $\nu = \frac{1}{k{\cdot}t_\text{delay}}$.
In addition, a set of $s$ connections are placed between ring neurons and random endpoints, where $w=0$.
These connections generate computational load in the code paths responsible for spike transmission and delivery, yet have no effect on the target synapses.
The cell model is generated randomly as a tree of given depth, which is done individually per cell to avoid early-stage optimizations. 
The neuron is then endowed with Hodgkin-Huxley dynamics \citep{HodgkinHuxley1952} on the soma region and a plain leak current on the remaining morphology. We call this neuron model \texttt{simple-branchy}. 
To consider the computational load of plasticity dynamics, we further implemented a variant of the benchmark where the synapses targeted by $w=0$ connections use an STDP model (see subsection \ref{ssec:stdp}).

\begin{table}[]
    \small
    \begin{center}
    \begin{tabular}{|p{5.5cm}|p{3.0cm}|p{3.0cm}|}
    \hline {Description} & {Symbol} & {Value}\\
	\hline
	\hline
    Number of CPU threads & {-} & $4$ -- $64$\\
	\hline
    Number of MPI ranks & {-} & $4$ -- $64$\\
	\hline
    Simulated time & {$t_\text{duration}$} & $200\unit{ms}$\\
	\hline
    Timestep & ${\Delta t}$ & $0.025\unit{ms}$\\
	\hline
    Synaptic delay & $t_\text{delay}$ & $5\unit{ms}$\\
	\hline
    Synaptic time constant & $\tau$ & $2\unit{ms}$\\
    \hline
    Decay of the STDP eligibility trace of presynaptic spikes & $\tau_\textrm{pre}$ & {$10.0\unit{ms}$}\\
    \hline
    Decay of the STDP eligibility trace of postsynaptic spikes & $\tau_\textrm{post}$ & {$10.0\unit{ms}$}\\
    \hline
    STDP strengthening amplitude & $A_\textrm{pre}$ & {$0.01\unit{\textmu{S}}$}\\
    \hline
    STDP weakening amplitude & $A_\textrm{post}$ & {$-0.01\unit{\textmu{S}}$}\\
    \hline
    Maximum synaptic weight for STDP & $w_\textrm{max}$ & {$10.0\unit{\textmu{S}}$}\\
	\hline
    Ring size & $k$ & $4$\\
	\hline
    Random connections per cell & $s$ & $1000$\\
	\hline
    \end{tabular}
    \caption{%
      {\small \textbf{Parameters for benchmarking large networks.} 
      Used for benchmarking large networks with busyring in Arbor and CoreNEURON (STDP parameters are only used in Arbor).
      For CPU threads and MPI ranks, all combinations of powers of $2$ from the given ranges were considered (the ranges were chosen according to the 64-core CPU of the HWS2 system).
    \label{tab:parameters_benchmarking_very_large}}}
    \end{center}
\end{table}

 

We measured the resulting wallclock times on our hardware system `HWS2' that we already used in subsection \ref{ssec:benchmarking}, using the parameter values from Table~\ref{tab:parameters_benchmarking_very_large}.
Next, we determined the optimal runtime on this multi-core CPU system by selecting the best performance across the given ranges of MPI ranks and CPU threads, and subsequently compared the results for CoreNEURON, Arbor, and Arbor with STDP synapses (Fig.~\ref{fig:benchmarking_runtime_very_large}a-c). From this, we observe a tremendous speedup provided by Arbor, even if STDP dynamics are considered.
Fig.~\ref{fig:benchmarking_runtime_very_large}d shows that the speedup is also maintained across network sizes, where both CoreNEURON and Arbor scale almost linearly. 
Furthermore, disentangling the setup and state propagation phases of the simulations, we find almost linear scaling for these particular measures as well (Fig.~\ref{fig:benchmarking_runtime_very_large}f).
Exact runtime values, also for different dendritic tree depths, are provided in Table~\ref{tab:runtime_very_large}. Note that Arbor's performance may be increased even further by appropriately adjusting the \texttt{cpu\_group\_size} setting for each particular case (here, all results are for \texttt{cpu\_group\_size=1}).

Regarding the memory use, we also observe a much better efficiency of Arbor as compared to CoreNEURON, while optimal results are achieved for the lowest rank and thread numbers (Supplementary Fig.~S14). Again both simulators scale almost linearly (see Supplementary Table~S1 for more details).

When additionally utilizing the existing GPU in the HWS2 setup described above, further speed gains can be achieved. 
In general, due to their large numbers of cores, it is expected that GPUs will accelerate simulations of large networks, which is supported by our results.
Comparing Fig.~\ref{fig:benchmarking_runtime_very_large}f and Fig.~\ref{fig:benchmarking_runtime_very_large}g, we see that for large networks (here: 32768 neurons), especially if they also feature plasticity dynamics, the GPU brings benefits for both setup and propagation phase (see dark gray bars in Fig.~\ref{fig:benchmarking_runtime_very_large}g; also cf. Tables~\ref{tab:runtime_very_large} and \ref{tab:runtime_very_large_gpu}).
However, for smaller networks of 1024 neurons, using the GPU can even slow down the simulation. 
For the exact runtime values with GPU, including such for different dendritic tree depths, see Table~\ref{tab:runtime_very_large_gpu}. Note that for CoreNEURON, none of the simulations finished because in all cases the compute system ran out of memory.


\begin{table}[!ht]
\small
\begin{center}
\begin{tabular}{|p{1.2cm}|p{0.8cm}|p{1.5cm}|p{2.0cm}|p{2.0cm}|p{2.0cm}|p{2.0cm}|}
    \hline {Number of cells} & {Tree depth} & {Number of synapses} & {Number of compartments} & {Runtime of CoreNEURON (s)} & {Runtime of Arbor (s)} & {Runtime of Arbor with STDP (s)}\\
	\hline
	\hline
	\multirow{4}*{\parbox{1.5cm}{1024}} & {0} & {1.03M} & {1024} & {2.83 (0.75+2.07)} & {0.23 (0.04+0.19)} & {0.73 (0.05+0.68)}\\
	\cline{2-7}
	{} & {2} & {1.03M} & {46116} & {2.81 (0.78+2.04)} & {0.32 (0.04+0.28)} & {0.84 (0.05+0.79)}\\
	\cline{2-7}
	{} & {10} & {1.03M} & {1672700} & {6.23 (0.65+5.58)} & {2.67 (0.19+2.48)} & {3.22 (0.21+3.01)}\\
	\hline
	\multirow{4}*{\parbox{1.5cm}{16384}} & {0} & {16.41M} & {16384} & {51.49 (5.28+46.21)} & {3.24 (0.37+2.86)} & {11.36 (0.75+10.61)}\\
	\cline{2-7}
	{} & {2} & {16.41M} & {737408} & {53.52 (5.32+48.20)} & {4.64 (0.41+4.23)} & {13.18 (0.81+12.37)}\\
	\cline{2-7}
	{} & {10} & {16.41M} & {26616896} & {163.11 (7.47+155.64)} & {43.07 (2.35+40.72)} & {52.48 (2.58+49.90)}\\
    \hline
	\multirow{4}*{\parbox{1.5cm}{32768}} & {0} & {32.81M} & {32768} & {103.61 (9.57+94.04)} & {6.11 (0.80+5.31)} & {22.49 (1.48+21.00)}\\
	\cline{2-7}
	{} & {2} & {32.81M} & {1475268} & {106.17 (9.90+96.27)} & {8.68 (0.82+7.86)} & {26.16 (1.72+24.45)}\\
	\cline{2-7}
	{} & {10} & {32.81M} & {52957704} & {d.n.f.} & {85.75 (4.65+81.09)} & {104.66 (5.05+99.60)}\\
    \hline
\end{tabular}
\caption{\small \textbf{Wallclock time measurements for busyring benchmark.} 
Total runtime results are provided as reported by Arbor and CoreNEURON. The shares of the setup and state propagation phases are given in brackets, respectively. Results are collected with the HWS2 system (AMD Ryzen Threadripper PRO 5995WX CPU with 64 cores, $8\times32\unit{GB}$ DDR4-RAM, without GPU). All values are averaged over $10$ trials, with coefficient of variation $\mathrm{CV} < 0.06$ in all cases. Arbor with SIMD. In CoreNEURON, the most extensive simulation did not finish (d.n.f.) due to exceeded memory. See Table~\ref{tab:runtime_very_large_gpu} for results with GPU.\label{tab:runtime_very_large}}
\end{center}
\end{table}

\begin{table}[!ht]
\small
\begin{center}
\begin{tabular}{|p{1.2cm}|p{0.8cm}|p{1.5cm}|p{2.0cm}|p{2.0cm}|p{2.0cm}|}
    \hline {Number of cells} & {Tree depth} & {Number of synapses} & {Number of compartments} & {Runtime of Arbor (s)} & {Runtime of Arbor with STDP (s)}\\
	\hline
	\hline
	\multirow{4}*{\parbox{1.5cm}{1024}} & {0} & {1.03M} & {1024} & {0.32 (0.02+0.30)} & {0.81 (0.03+0.78)}\\
	\cline{2-6}
	{} & {2} & {1.03M} & {46116} & {0.44 (0.02+0.41)} & {1.03 (0.03+1.00)}\\
	\cline{2-6}
	{} & {10} & {1.03M} & {1672700} & {3.13 (0.14+2.99)} & {3.65 (0.15+3.50)}\\
	\hline
	\multirow{4}*{\parbox{1.5cm}{16384}} & {0} & {16.41M} & {16384} & {3.10 (0.29+2.81)} & {10.94 (0.50+10.44)}\\
	\cline{2-6}
	{} & {2} & {16.41M} & {737408} & {4.46 (0.32+4.15)} & {12.60 (0.56+12.04)}\\
	\cline{2-6}
	{} & {10} & {16.41M} & {26616896} & {43.51 (3.07+40.45)} & {52.38 (3.33+49.05)}\\
    \hline
	\multirow{4}*{\parbox{1.5cm}{32768}} & {0} & {32.81M} & {32768} & {5.85 (0.55+5.30)} & {21.51 (1.05+20.47)}\\
	\cline{2-6}
	{} & {2} & {32.81M} & {1475268} & {8.48 (0.70+7.78)} & {24.84 (1.13+23.71)}\\
	\cline{2-6}
	{} & {10} & {32.81M} & {52957704} & {86.62 (6.18+80.44)} & {104.46 (6.71+97.75)}\\
    \hline
\end{tabular}
\caption{\small \textbf{Wallclock time measurements for busyring benchmark with GPU.} 
Total runtime results are provided as reported by Arbor. The shares of the setup and state propagation phases are given in brackets, respectively. Results are collected with the HWS2 system (AMD Ryzen Threadripper PRO 5995WX CPU with 64 cores, $8\times32\unit{GB}$ DDR4-RAM, with NVIDIA T1000 8GB GPU). All values are averaged over $10$ trials, with coefficient of variation $\mathrm{CV} < 0.06$ in all cases. Arbor with SIMD. In CoreNEURON, due to exceeded memory, none of the simulations finished. See Table~\ref{tab:runtime_very_large} for results without GPU.\label{tab:runtime_very_large_gpu}}
\end{center}
\end{table}

By choosing different values for $t_\text{delay}$, $k$, and $s$, the workload can be adapted to investigate the network behavior of different real-world models. 
Considering different cell models further allows the emulation of their computational workload in network models.
The relation of both network and cell parameters can then provide a holistic image of the performance of varying network models.
In addition to the \texttt{simple-branchy} cell with Hodgkin-Huxley dynamics that we considered above, Arbor's busyring implementation features another pre-configured cell model called \texttt{complex}. This model comprises a set of eight channel types on the soma, including kinetic schemes and calcium concentration models, as well as five channel types on the remaining morphology. In this way, the model approximates a cell from the mouse visual cortex \cite{allen-db}.
A runtime comparison between \texttt{complex} and \texttt{simple-branchy} is shown in  Supplementary Table~S2. 

The scaling behavior of Arbor enables the simulation of even larger and more complex network models than the ones we considered so far in this study. As an outlook, we show in Supplementary Fig.~S15 first results of strong scaling on state-of-the-art supercomputing systems at JSC (the preview system JEDI and the current flagship system JUWELS) with an extremely large network of $10^6$ cells of the \texttt{complex} type (nevertheless, without plasticity dynamics).
The parameters for both systems are given in Supplementary Table~S3.
As the NVIDIA H100 GPU model used in JEDI offers roughly twice the memory bandwidth and more than twice the floating-point performance of the A100 of JUWELS, we compare one JEDI node to two nodes of JUWELS.
Within the strong scaling range over an eightfold increase of nodes in the simulation, we observe a scaling efficiency $\epsilon$ of well over 80\%. The scaling efficiency is defined by
\begin{align*}
    \epsilon = \frac{T(n_0)}{T(n)\cdot n},
\end{align*}
where $T(n)$ is the wallclock time measured in the strong scaling series with $n$ nodes and reference $n_0=4$.

\section{Availability and Future Directions}

\subsection{Discussion and outlook}

In this work, we have aimed to demonstrate the versatility of the extended Arbor simulator in modeling synaptic plasticity mechanisms within large neuronal networks. Specifically, we considered Arbor implementations of homosynaptic, homeostatic, and heterosynaptic plasticity mechanisms in different setups.
In subsections \ref{ssec:stdp} \& \ref{ssec:homeostasis}, we presented plasticity rules that can be considered a basis for further spike-based mechanisms of synaptic plasticity. 
In subsections \ref{ssec:ca_plast_graupner_brunel}--\ref{ssec:stc_synapse_and_network}, we considered three different calcium-based rules in different scenarios. The results in subsection \ref{ssec:ca_plast_graupner_brunel} constitute a reproduction of a widely-used calcium model that was directly fitted to experimental data \citep{GraupnerBrunel2012}. The results in subsection \ref{ssec:heterosyn_ca_plast} show how a model of calcium diffusion along dendrites can be employed to simulate heterosynaptic plasticity (cf. \citep{HirataniFukai2017,Shafiee2024}).
In subsection \ref{ssec:stc_synapse_and_network}, we used a calcium model as basis for a more complex model that captures early- and heterosynaptic late-phase plasticity \citep{Li2016,LuboeinskiTetzlaff2021,LuboeinskiTetzlaff2024}. 
Finally, we moved to the network level (subsections \ref{ssec:stc_synapse_and_network} \& \ref{ssec:stc_morpho_neuron_network}), where we built on the complex two-phase learning rule introduced before. Here, we first reproduced previous results on memory recall \citep{LuboeinskiTetzlaff2021} using single-compartment neurons, and then extended the neurons by additional morphological segments accounting for dendritic structure. 
By this, we could demonstrate how (Plastic) Arbor serves to seamlessly gather computational insights into the impact of morphological neuron structures in large networks. 
In particular, our results from the multi-compartment model provide new insights showing that large dendritic structures can have a deteriorating impact on memory function at the network level, and that the PRP transport velocity in these structures might only play a minor role (Fig.~\ref{fig:stc_multicomp_network}e--h). Furthermore, a large cell diameter can have a converse effect, yielding enhanced memory recall (Fig.~\ref{fig:stc_multicomp_network}i--l).

In the following, we will briefly discuss the pros and cons of the most prominent neural network simulators, and compare their target use cases with Arbor.
With its first version released in the 1980s and widespread usage, NEURON has significantly advanced the understanding of the brain by facilitating the computational study of complex neuronal processes \citep{CarnevaleHines2006}. Nevertheless, as mentioned above, its engine under the hood is outdated when it comes to high-performance computing, especially, for network simulations involving detailed neuron models. While CoreNEURON \citep{Kumbhar2019, Awile2022} constitutes an approach to address this, 
it has restricted flexibility and usability due to its dependence on the NEURON environment, and it lacks support for certain important features of NMODL \citep{nrn-docs-coreneuron-compatibility}. 
Further, GENESIS (GEneral NEural SImulation System) \citep{BowerBeeman2003} is a simulator that has also been used widely for several decades, offering a platform for multi-scale simulations with particular focus on detailed electrical and chemical interactions. 
While GENESIS is very limited with respect to modern hardware backends, the MOOSE simulator is developed as its modern successor \citep{Ray2008}. 
Regarding alternatives that are more focused on point-neuron simulations, NEST (NEural Simulation Tool) is a widely used simulator designed for large-scale simulations of spiking neural networks. 
It is particularly known for its scalability and efficiency in simulating large-scale networks. However, its lack of support for multi-compartment neurons has been one of the initial reasons to develop Arbor, which was therefore originally named `NestMC'. 
Brian 2 is another widely used network simulator that enables the definition of models directly via differential equations \citep{Stimberg2019}. Furthermore, Brian 2 enables high flexibility by generating an intermediate abstract code, processed by so-called device interfaces. This allows to seamlessly exchange the underlying numerical backend architecture. While Brian 2 does not yet come with comprehensive support for multi-compartment neurons, a new extension called Dendrify specifically focuses on detailed dendritic morphology and may provide a valuable tool for the investigation of dendritic processing \citep{Pagkalos2023}.
CARLsim \citep{Niedermeier2022} is another framework for the parallelized simulation of large-scale spiking neural networks, but it has been optimized for real-time simulations and neuromorphic hardware implementation. Accordingly, the single-compartment Izhikevich neuron is the most biologically realistic neuron model that is supported by CARLsim.
Finally, the recently released simulator EDEN \citep{Panagiotou2022} appears to offer high flexibility by supporting NeuroML model descriptions, however, it still lacks support for GPU backends. As opposed to that, the equally new simulator NEST GPU (previously named NeuronGPU) \citep{Golosio2021,Golosio2023} is optimized for neural network simulations on GPU, however, it does not support multi-compartment models. Similarly, the GeNN simulator also offers highly optimized GPU simulation for networks of point neurons \citep{Knight2021}.

In summary, each of the existing simulators comes with its own strengths and weaknesses (also see \citep{TikidjiHamburyan2017,Kulkarni2021,Wang2023}). The decision to use one specific simulator should depend on the particular needs of a research project, the available compute resources, as well as the expertise of the involved researchers. For example, Brian 2 can be considered quite user-friendly for quickly setting up networks of point neurons, NEST exhibits unique performance for very large network simulations, and for NEURON there are many existing implementations of morphological neuron models. 
Arbor, finally, was designed to easily define networks of morphological neurons and then to map the internal modeling primitives to available compute resources in an optimized manner. As we have shown, this approach enables very good runtime efficiency and makes Arbor an attractive option, particularly, for researchers who rely on high-performance computing for their simulations with networks of morphologically detailed neurons and existing NEURON models.

We have cross-validated all of our presented Arbor implementations either with Brian 2 \citep{Stimberg2019} or with one of multiple stand-alone simulators that were custom-developed for previous studies \citep{GraupnerBrunel2012,StandAloneCode,HeterosynapticCalciumCode}. 
At the single-synapse level, we did not find any significant differences between the results of the simulators (Figs.~\ref{fig:stdp_homeostasis_calcium}--\ref{fig:stc_synapse_and_singlecomp_network}; Supplementary Figs.~S1--S3). However, this does not preclude that certain differences may arise in other simulation paradigms, where the validity of the particular simulators that are used should be carefully evaluated (also cf. \cite{Plesser2024}).
Furthermore, differences in the neuron model and in the numerical methods, as well as complex network effects, give rise to certain quantitative deviations in memory recall performance at the circuit level, while the qualitative behavior remained the same (Supplementary Fig.~S10; also cf. subsection \ref{ssec:stc_synapse_and_network} and Supplementary Fig.~S4). 

Regarding the use of compute resources for a large network with plastic connections, we found that Arbor performs well both in terms of runtime and memory use. Compared to optimized point-neuron simulators, Arbor only uses slightly more resources when computing on CPU, and can even outperform those simulators when using its capability to compute on GPU (Fig.~\ref{fig:benchmarking_memory_consolidation}).
However, Arbor particularly shines when it comes to simulating networks of multi-compartment neurons, which necessitates almost no additional cost compared to single-compartment neurons (shown by the runtime per spike, Supplementary Fig.~S12). This is not entirely unexpected, since Arbor has been designed particularly for networks of multi-compartment neurons. Furthermore, we found that Arbor outperforms its main competitor in this regard, CoreNEURON, in terms of runtime and memory use across all considered network sizes, even with an additional plasticity workload (Tables~\ref{tab:runtime_very_large} \& \ref{tab:runtime_very_large_gpu}, Supplementary Table S2).

The benchmarking results notwithstanding, there may be several ways to even further improve the performance of Arbor.
First, to compute the dynamics of given models, Arbor uses implicit solving methods, which have the advantage of being stable but come at the cost of runtime performance. In specific cases, these methods may be replaced by faster explicit algorithms. Kobayashi et al. \citep{Kobayashi2021}, for instance, have shown how an explicit method with adaptive time steps and second-order accuracy can serve to avoid heavy memory access, which can be helpful particularly when using GPUs. 
Second, the introduction of exact point neurons may eliminate the need to simulate a spatial neuron model if this is not needed. Note that meanwhile, a LIF neuron feature has been added to Arbor, although still being in a test stage.

By the final model simulation results that we have presented in Fig.~\ref{fig:stc_multicomp_network}, we could gain first new insights into the functional interplay between, on the one hand, network parameters such as the size of a stimulated pattern, and on the other hand, PRP diffusion within neurons. Specifically, the results suggest that the functionality of pattern recall is not very sensitive to the diffusion speed. Nevertheless, if the diffusion is too slow, as expected, all memory functionality will vanish, which also occurs if the pattern size is too small. Moreover, if we increase the size of the dendritic branches (in a range that roughly relates to lengths of main branches in cortical pyramidal cells \citep{CarribaDavies2017,vanAerde2015,CarnevaleHines2006}), the memory recall capability is impaired (Fig.~\ref{fig:stc_multicomp_network}f--h), indicating that additional mechanisms may be needed to obtain a stable memory system. Somehow in contrast to that, a larger cell and dendrite diameter improves memory recall (Fig.~\ref{fig:stc_multicomp_network}j--l). 
By targeting the impact that PRP diffusion within neurons has on the dynamics of large networks, our findings may complement the picture that other studies have drawn of the functional role of spatial PRP dynamics within neurons \cite{Sajikumar2007,ODonnellSejnowski2014,Kastellakis2016,Fonkeu2019,Sartori2020}.

In the future, most importantly, the extended Arbor simulation framework can enable researchers to conduct studies that examine the interplay between neuron-internal, synaptic, and network dynamics. These may, for instance, be related to memory consolidation \citep{LuboeinskiTetzlaff2021,LuboeinskiTetzlaff2024,ZenkeLaborieux2024} or working memory \citep{Eriksson2015,Masse2020}. Moreover, Arbor allows the implementation of models that include changes in the connectivity structure of networks. Using this together with the tools that we have presented here will enable researchers to also study the interactions between structural and functional plasticity processes at the network level (cf.~\citep{ODonnell2011,Fauth2017,Gallinaro2022,Lu2023,Kaster2024}).
Another important goal will be to add the models that have so far been implemented in Arbor to open databases, such as the Open Source Brain repository or ModelDB, for broader adoption and community-driven validation.
It would also be intriguing to conduct collaborative studies where Arbor is used to simulate electrophysiological and network properties, while at the same time using calcium imaging experiments to validate diffusion dynamics. This could serve to validate models as given in section \ref{ssec:heterosyn_ca_plast} \cite{Shafiee2024}. Furthermore, spine density distributions derived from in vivo experiments \cite{Yang2009,Attardo2015} could be used, for example, to test diffusion models in more detail.
It should further be noted that by considering synaptic plasticity processes in neural networks, simulation software contributes essentially to the understanding of biological mechanisms as well as to the development of artificial intelligence applications. 
In this light, further applications of our extended version of Arbor may target, for example, paradigms of somato-dendritic mismatch error reduction \cite{Senn2023}, reservoir computing with heterogeneous networks \citep{PerezNieves2021,Morales2021}, or prototyping and benchmarking of neuromorphic algorithms (cf.~\citep{Kulkarni2021,Boahen2022,Khacef2023}). 

Due to its modern computing architecture and inherent support of multi-compartment neurons, the Arbor simulator constitutes an important tool for the  computational modeling of neuronal networks. Through the newly introduced extensions, enabling to simulate synaptic plasticity, we have increased the range of Arbor's use cases substantially. Furthermore, as an example, we have provided a first demonstration of how the extended framework enables to gain new insights into the functional impact of morphological neuronal structure at the network level.
With its newly extended functionality, the Arbor framework is able to power a great variety of future studies considering synaptic mechanisms and their interactions with neuronal morphologies, from single synapses to large networks, in a highly efficient manner.


\subsection{Data and code availability}
All data presented in this study can be reproduced using publicly available simulation code and analysis scripts, which are listed in the following.

The Arbor framework can be installed as described on \href{https://arbor-sim.org/}{https://arbor-sim.org}. The code can be retrieved, e.g., from \citep{ArborCoreCode}. Our simulation code for the different subsections is referenced in Table~\ref{tab:code_references} below. 
Note that since Arbor is still under development, some parts of our model implementations may be subject to changes when using later Arbor versions, however, we intend to keep the model implementations updated along with the Arbor core code.

Due to the ongoing development of the Arbor core code, we ran our extensive simulations with different Arbor versions. 
For the simulation results presented in subsections \ref{ssec:stdp}--\ref{ssec:ca_plast_graupner_brunel} as well as \ref{ssec:stc_synapse_and_network}--\ref{ssec:stc_morpho_neuron_network}, 
we used Arbor version \href{https://github.com/arbor-sim/arbor/commit/2f4c32598d37f9852978c76952b0a09aeb84385b}{0.9.1-dev-2f4c325} (for instructions on how to install this exact version, see the \texttt{README.md} file that comes with our model code \citep{ArborNetworkScCode}). For the simulation results of subsection \ref{ssec:heterosyn_ca_plast}, we used the development branch of pull request \#2226 with the state of commit \href{https://github.com/arbor-sim/arbor/pull/2226/commits/f0e456d631bf818eddee870167828a065dc4afa7}{f0e456d}, which meanwhile has been merged (note that at that state of commit, Arbor still used diffusion constants in non-standard units, scaled by a factor of $10^{-7}$).
For the benchmarking results in subsection \ref{ssec:benchmarking} we used Arbor version \href{https://github.com/arbor-sim/arbor/releases/tag/v0.10.0}{0.10.0}, and for those in subsection \ref{ssec:benchmarking_very_large} we used the state of commit \href{https://github.com/thorstenhater/arbor/tree/cebb65bf53f9b341d8a4fdd300f53e07002643dd}{cebb65b} of an external fork, which shall be merged upon publication of this article. 
Note that while all of the mentioned Arbor versions are by now slightly outdated, we ensured that our extensive software tests from \citep{ArborNetworkScCode,ArborNetworkMcCode} passed with the used versions including the release version \href{https://github.com/arbor-sim/arbor/releases/tag/v0.10.0}{0.10.0}, 
which strongly indicates that the simulation results will remain the same.

The scripts that we used to perform the comparisons across simulators in subsection \ref{ssec:stc_synapse_and_network} can be found here: \href{https://github.com/jlubo/simulator_comparison}{https://github.com/jlubo/simulator\_comparison}.

\begin{table}[!ht]
\small
\begin{center}
\begin{tabular}{|p{0.5cm}|p{4.2cm}|p{3.9cm}|p{3.9cm}|}
    \hline {Sec\-tion} & {Model description} & {Arbor implementation} & {Reference implementation(s)} \\
    \hline
    \hline {\ref{ssec:stdp}} & {Spike-timing-dependent plasticity} & {\citep{FippaCode} (subdirectory \texttt{STDP/})} & {\citep{FippaCode} (subdirectory \texttt{STDP/}) 
    }\\
	\hline {\ref{ssec:homeostasis}} & {Spike-driven homeostatic plasticity} & \citep{FippaCode} (subdirectory \texttt{spike\_based\_homeostasis/}) & {\citep{FippaCode} (subdirectory \texttt{spike\_based\_homeostasis/})
    }\\
	\hline{\ref{ssec:ca_plast_graupner_brunel}} & {Calcium-based synaptic plasticity} & \citep{ArborCoreCode} (Arbor main repository, \texttt{calcium\_stdp.py}) & {\citep{GraupnerBrunelCode}}\\
	\hline {\ref{ssec:heterosyn_ca_plast}} & {Heterosynaptic calcium-based plasticity in dendrites} & {\citep{HeterosynapticCalciumCode} (\texttt{Arbor\_diff.py})} & {\citep{HeterosynapticCalciumCode} (\texttt{Custom.py}) }\\
	\hline {\ref{ssec:stc_synapse_and_network}} & {Synaptic tagging and capture, in individual synapses and in networks of single-compartment neurons} & \citep{Arbor2N1SCode} (reduced code for single synapses) and {\citep{ArborNetworkScCode}} & {\citep{StandAloneCode}; \citep{Brian2N1SCode} (reduced code for single synapses) and \citep{BrianNetworkCode}
    }\\
	\hline {\ref{ssec:stc_morpho_neuron_network}} & {Synaptic memory consolidation in networks of morphological neurons} & {\citep{ArborNetworkMcCode}} & {- (novel results)}\\
	\hline {\ref{ssec:benchmarking}} & {Network models from \ref{ssec:stc_synapse_and_network} and \ref{ssec:stc_morpho_neuron_network}} & {\citep{ArborNetworkScCode}; \citep{ArborNetworkMcCode}} & {\citep{StandAloneCode}; \citep{BrianNetworkCode}}\\
	\hline {\ref{ssec:benchmarking_very_large}} & {Busyring network benchmark} & {\citep{busyringArbor}} & {\citep{busyringCoreNeuron}}\\
	\hline
\end{tabular}
\caption{\small Overview of the simulation code used to perform the simulations presented in this article.\label{tab:code_references}}
\end{center}
\end{table}

\clearpage


\addcontentsline{toc}{section}{Acknowledgments}
\section*{Acknowledgments}

We would like to thank the other Arbor developers for their support and for their dedication to continuously improve the software. We would further like to thank Silvio Rizzoli, Arash Golmohammadi, and Marcel Stimberg for helpful discussions.

\addcontentsline{toc}{section}{Financial Disclosure}
\section*{Financial Disclosure}
For this work, CT received funding from the German Research Foundation (Deutsche Forschungsgemeinschaft, DFG) through grants SFB1286 (C01, Z01) and TE 1172/7-1, as well as from European Commission Horizon 2020 through grant no. 899265
(ADOPD). Further, CT and SSc received funding from European Commission Horizon 2020 through grant no. 945539 (HBP SGA3). The funders had no role in study design, data collection and analysis, decision to publish, or preparation of the manuscript.

\addcontentsline{toc}{section}{Author Contributions}
\section*{Author Contributions}

Conceptualization: JL, SSc, CT;
Data curation: JL, SSc, SSh;
Formal Analysis: JL, SSc, SSh;
Funding acquisition: SSc, CT;
Investigation: JL, SSc, SSh, CT;
Methodology: JL, SSc, SSh, TH, CT;
Project administration: CT;
Resources: TH, CT;
Software (Arbor core): JL, SSc, TH, FB (with larger contributions by TH and FB);
Software (sections~\ref{ssec:stdp}--\ref{ssec:homeostasis}): JL, SSc;
Software (section~\ref{ssec:ca_plast_graupner_brunel}): SSc, FB;
Software (section~\ref{ssec:heterosyn_ca_plast}): SSh;
Software (sections~\ref{ssec:stc_synapse_and_network}--\ref{ssec:benchmarking}): JL;
Software (section~\ref{ssec:benchmarking_very_large}): JL and TH;
Supervision: JL, SSc, CT;
Validation: JL, SSc, SSh, TH, FB; 
Visualization: JL, SSc, SSh, TH;
Writing -- original draft: JL, SSc, SSh, TH, FB;
Writing -- review \& editing: JL, SSc, CT.

\clearpage
\newpage
\addcontentsline{toc}{section}{References}
\bibliography{references}

\clearpage
\addcontentsline{toc}{section}{List of Legends of the Supplementary Information}
\section*{List of Legends of the Supplementary Information}

\noindent {\small Figure S1: \textbf{More details on classical spike-timing dependent plasticity (STDP) and spike-driven homeostasis.} Related to \red{Fig.~2} in the main article. Arbor implementations (in lighter blue) are cross-validated by comparison to Brian 2 (in orange) or theory.
In (a--d), two Poisson spike sources stimulate an inhibitory and an excitatory synapse connecting to a single neuron (spikes are shown in red and blue, respectively). The excitatory connection undergoes STDP. \textbf{(a)} Membrane potential of the neuron (goodness of fit between the curves: $\mathrm{CV} = 0.923$, $\mathrm{RMSE} = 1.265\unit{mV}$). \textbf{(b)} Conductance of the excitatory synapse ($\mathrm{CV} = 0.996$, $\mathrm{RMSE} = 0.486\unit{{\textmu}S}$). \textbf{(c)} Conductance of the inhibitory synapse ($\mathrm{CV} = 0.997$, $\mathrm{RMSE} = 0.148\unit{{\textmu}S}$). \textbf{(d)} Spike time mismatch measured by $\left(t^\text{Arbor}-t^\text{Brian}\right)/t^\text{Brian} \cdot 100$, where $t^\text{Arbor}$ and $t^\text{Brian}$ are the postsynaptic spike times obtained from the Arbor and Brian 2 simulations, respectively. The result indicates that the difference in spike timing due to different implementations of the Poisson process is below $0.1\%$. \textbf{(e)} Classical STDP curve, compared with theoretical expectation $A_\textrm{pre} \cdot \exp(-\Delta t / \tau_\textrm{pre})$ for $\Delta t>0$ and $ A_\textrm{post} \cdot \exp(\Delta t / \tau_\textrm{post})$ otherwise ($\mathrm{CV} > 0.999$, $\mathrm{RMSE} = 0.001\unit{{\textmu}S}$). \textbf{(f)} As opposed to \red{Fig.~2i} in the main article, the resulting firing rate of the neuron in the absence of homeostatic plasticity is shown ($\mathrm{CV} = 0.972$, $\mathrm{RMSE} = 1.674\unit{Hz}$).}
\\

\noindent {\small Figure S2: \textbf{Basic early- and late-phase plasticity dynamics with synaptic tagging and capture (STC), cross-validated with stand-alone simulator.} Also note the the cross-validation with the stand-alone simulator in the main article (\red{Fig.~4}) and with Brian 2 in \red{Fig.~S3}.  
\textbf{(a)} Averaged noisy early-phase synaptic weight (see \red{Eq.~17}) in the main article). Stimulating spikes reach the synapse at pre-defined times (indicated by bold gray arrows). Goodness of fit between the mean curves: $\mathrm{CV} = 0.999$, $\mathrm{RMSE} = 0.040\unit{mV}$. \textbf{(b)} Limit cases of early- and late-phase synaptic weight (see \red{Eq.~17} and \red{Eq.~20} in the main article). The presynaptic neuron is stimulated with a strong current to spike at maximal rate (duration of the stimulation indicated by gray bar). The late-phase weight has been shifted for graphical reasons (also cf. \red{Eq.~20}; early phase: $\mathrm{CV} = 0.201$, $\mathrm{RMSE} = 0.221\unit{mV}$; late phase: $\mathrm{CV} > 0.999$, $\mathrm{RMSE} = 0.055\unit{mV}$). 
\textbf{(c)} Standard deviation of the noisy early-phase synaptic weight ($\mathrm{RMSE} = 0.067\unit{mV}$), and
\textbf{(d)} standard deviation of early- and late-phase synaptic weight (early phase: $\mathrm{RMSE} = 0.133\unit{mV}$; late phase: $\mathrm{RMSE} = 0.004\unit{mV}$), demonstrating the matching of the stochastic properties of the two solvers.
\textbf{(e)} Postsynaptic calcium concentration, which successively crosses the thresholds for depression (LTD) and potentiation (LTP) (cf. Eq.~\red{Eq.~19} in the main article; $\mathrm{CV} > 0.999$, $\mathrm{RMSE} = 0.065$). 
\textbf{(f)} The postsynaptic PRP concentration rises until it reaches its maximum due to the continued stimulation (cf. \red{Eq.~21} in the main article; $\mathrm{CV} = 0.998$, $\mathrm{RMSE} = 0.107\unit{{\textmu}mol/l}$). 
\textbf{(g)} Membrane potential of the postsynaptic neuron ($\mathrm{CV} > 0.999$, $\mathrm{RMSE} = 0.151\unit{mV}$). 
Continuous lines (specified in the legends) represent the dynamics simulated in Arbor. Darker, dashed lines represent the results of the stand-alone simulator \citep{StandAloneCode}. Fine, dotted lines represent the baseline level of the respective quantity. 
Basic early-phase plasticity dynamics (a,c,e,g): average across $10$ batches, each consisting of $100$ trials. Basic late-phase plasticity dynamics (b,d,f): average across $10$ batches, each consisting of $10$ trials. The noise seeds were drawn independently for each trial. Error bands represent the standard error of the mean (often too small to be visible).}
\\

\noindent {\small Figure S3: \textbf{Basic early- and late-phase plasticity dynamics with synaptic tagging and capture (STC), cross-validated with the Brian 2 simulator.} Also see the the cross-validation with the stand-alone simulator in \red{Fig. S2} and in the main article (\red{Fig.~4}).  
\textbf{(a)} Averaged noisy early-phase synaptic weight (see \red{Eq.~17}) in the main article). Stimulating spikes reach the synapse at pre-defined times (indicated by bold gray arrows). \textbf{(b)} Limit cases of early- and late-phase synaptic weight (see \red{Eq.~17} and \red{Eq.~20} in the main article). The presynaptic neuron is stimulated with a strong current to spike at maximal rate (duration of the stimulation indicated by gray bar). The late-phase weight has been shifted for graphical reasons (also cf. \red{Eq.~20}). 
\textbf{(c)} Standard deviation of the noisy early-phase synaptic weight, and
\textbf{(d)} standard deviation of early- and late-phase synaptic weight, demonstrating the matching of the stochastic properties of the two solvers.
\textbf{(e)} Postsynaptic calcium concentration, which successively crosses the thresholds for depression (LTD) and potentiation (LTP) (cf. Eq.~\red{Eq.~19} in the main article). \textbf{(f)} The postsynaptic PRP concentration rises until it reaches its maximum due to the continued stimulation (cf. \red{Eq.~21} in the main article). \textbf{(g)} Membrane potential of the postsynaptic neuron. 
Continuous lines (specified in the legends) represent the dynamics simulated in Arbor. Darker, dashed lines represent the results of \citep{Brian2N1SCode}, using Brian 2 with \texttt{cpp\_standalone} device \citep{Stimberg2019}. Fine, dotted lines represent the baseline level of the respective quantity. 
Basic early-phase plasticity dynamics (a,c,e,g): average across $10$ batches, each consisting of $100$ trials. Basic late-phase plasticity dynamics (b,d,f): average across $10$ batches, each consisting of $10$ trials. The noise seeds were drawn independently for each trial. Error bands represent the standard error of the mean (often too small to be visible).}
\\

\noindent {\small Figure S4: \textbf{Elementary motifs of spike transmission in a pre-defined sparsely connected network, cross-validated with stand-alone simulator}. Continuous lines (specified in the legends) represent the dynamics simulated in Arbor. Darker dashed lines represent the results of the stand-alone simulator \citep{StandAloneCode}. Stochastic variables in the model have been replaced by deterministic mean dynamics ($\sigma_\textrm{pl}=0$). The connectivity matrix \texttt{connections\_default.txt} from \citep{StandAloneCode} is used. \textbf{(a)} Response of neuron $68$ to a spike in neuron $6$ (excitatory$\rightarrow$excitatory); \textbf{(b)} response of neuron $1760$ to a spike in neuron $6$ (excitatory$\rightarrow$inhibitory); \textbf{(c)} response of neuron $17$ to a spike in neuron $1615$ (inhibitory$\rightarrow$excitatory); \textbf{(d)} response of neuron $1690$ to a spike in neuron $1615$ (inhibitory$\rightarrow$inhibitory).}
\\

\noindent {\small Figure S5: \textbf{UML sequence diagram for the single-compartment synaptic tagging and capture model.} The diagram describes the technical specification that is necessary to implement the model from \citep{LuboeinskiTetzlaff2021}. Essentially, the spiking neuron model produces pre- and postsynaptic spikes that give rise to calcium-based early-phase plasticity, which then elicits synaptic tagging, synthesis of plasticity-related products (PRPs), and late-phase plasticity.}
\\

\noindent {\small Figure S6: \textbf{Impact of classical stimulation protocols at a single synapse.} Different types of long-term synaptic plasticity are induced depending on the stimulation protocol \citep{Sajikumar2005,Li2016}: \textbf{(a)} late-phase potentiation through strong tetanic (STET) stimulation, \textbf{(b)} early-phase potentiation through weak tetanic (WTET) stimulation, \textbf{(c)} late-phase depression through strong low-frequency (SLFS) stimulation, \textbf{(d)} early-phase depression through weak low-frequency (WLFS) stimulation. See \citep{LuboeinskiTetzlaff2021} for details of the protocols. All protocols affect the early-phase weight (dark red lines) and lead to the crossing of the tag threshold (dotted red lines), whereas only the `strong' protocols lead to the crossing of the PRP synthesis threshold (dashed green lines), thereby enabling changes in late-phase weight (blue lines). The total synaptic weight (orange lines) is the sum of early- and late-phase weight. These results were obtained with Arbor using the code from \citep{Arbor2N1SCode} (which has the same basis as our network code \citep{ArborNetworkScCode,ArborNetworkMcCode}). In addition, for comparison, the total synaptic weight obtained from point-neuron simulations with the stand-alone simulator \citep{StandAloneCode} is shown in grated dark shading (overlapping with the total weight from Arbor simulations). Average over $100$ trials; sampling rate: $30/\text{min}$. Compartment measures in Arbor as detailed in \red{Table~5} in the main article. Error bands represent the relative standard deviation of early-phase, late-phase, and total synaptic weight. The late-phase weight has further been shifted for graphical reasons (cf. \red{Eq.~20} in the main article).}
\\

\noindent {\small Figure S7: \textbf{Formal characteristics of the diffusion along a dendritic cable in Arbor.} \textbf{(a)} A single compartment and its equivalent circuit in the cable model. \textbf{(b)} Important quantities of the diffusion mechanisms.}
\\

\noindent {\small Figure S8: \textbf{Diffusion of arbitrary particles along a dendrite in Arbor at different times (maximum diffusivity).} Example for paradigm of large dendrites, small cells, and maximum diffusivity $D_\textrm{p}=10^{-11}\unit{m\textsuperscript{2}/s}$ (cf. \red{Table 6} in the main article). \textbf{(a,b)} Plasticity-related product or protein (PRP) concentration in the center of the soma, and summed amount of the signal triggering PRP synthesis (SPS) across the whole cell. Furthermore, the amount of SPS in the whole cell is estimated from the concentration of SPS in the center of the soma, which is the actual driver of PRP synthesis (PS) in the model. As long as this quantity is above the PS threshold, PS happens (cf. \red{Eqs.~21 \&~23} in the main article). \textbf{(c,d)} PRP concentration at the basal end of the soma, and summed amount of PRP across the whole cell (under ongoing PS, converges towards $p_\textrm{max}$ times the total volume of the cell). \textbf{(e,f)} PRP concentration along apical dendrite. \textbf{(g,h)} PRP concentration along basal dendrite.}
\\

\noindent {\small Figure S9: \textbf{Diffusion of arbitrary particles along a dendrite in Arbor at different times (moderate diffusivity).} Example for paradigm of large dendrites, small cells, and moderate diffusivity $D_\textrm{p}=10^{-15}\unit{m\textsuperscript{2}/s}$ (cf. \red{Table 6} in the main article). \textbf{(a,b)} Plasticity-related product or protein (PRP) concentration in the center of the soma, and summed amount of the signal triggering PRP synthesis (SPS) across the whole cell. Furthermore, the amount of SPS in the whole cell is estimated from the concentration of SPS in the center of the soma, which is the actual driver of PRP synthesis (PS) in the model. As long as this quantity is above the PS threshold, PS happens (cf. \red{Eqs.~21 \&~23} in the main article). \textbf{(c,d)} PRP concentration at the basal end of the soma, and summed amount of PRP across the whole cell (under ongoing PS, converges towards $p_\textrm{max}$ times the total volume of the cell). \textbf{(e,f)} PRP concentration along apical dendrite. \textbf{(g,h)} PRP concentration along basal dendrite.}
\\

\noindent {\small Figure S10: \textbf{Memory recall in recurrent networks of single-compartment neurons and point neurons, after learning and after consolidation.} Plots show the pattern completion, measured by coefficient $Q$ (see \red{Eq.~22} in the main article) for a stimulated subset of varied size (a varied number of neurons are stimulated for learning/recall). Values at $10\unit{s}$ are shown in red and values at $8\unit{h}$ in blue (after learning and after consolidation, respectively). Value $Q>0$ indicates the successful recall of a memory representation. Results in \textbf{(a)} obtained with Arbor, in \textbf{(b)} from the custom stand-alone simulator \cite{StandAloneCode}, and in \textbf{(c)} from Brian 2 with \texttt{cpp\_standalone} device \citep{BrianNetworkCode} (no data for $8\unit{h}$ due to missing fast-forward computation implementation). Data averaged over $100$ network realizations (unlike in \cite{LuboeinskiTetzlaff2021}). Error bars represent the 95\% confidence interval.}
\\

\noindent {\small Figure S11: \textbf{Memory recall in a recurrent neural network after learning and after consolidation.} 
Analogous to \red{Fig.~5} in the main article, a measure of mutual information is considered here. Results obtained with Arbor for networks of different kinds of multi-compartment neurons, demonstrating the impact of different values of the PRP diffusivity $D_\textrm{p}$ on memory consolidation. Networks consist of `small' cells (diameter of $6\unit{{\textmu}m}$) or of `large' cells (diameter of $12\unit{{\textmu}m}$), with either small or large dendrites (in which cases each neuron comprises in total $31$ or $48$ compartments, respectively). The diameter and length values are given in \red{Table~6} in the main article. \textbf{(a,b)} Illustrations of used cell structures, generated using \texttt{Arbor~GUI} \citep{Arbor-GUI-0-8-0}. Each segment is represented by a different color. A segment can consist of a multitude of compartments. Overlaid with illustrations of more realistic neuron structures that would have roughly similar functional properties. \textbf{(a)} Small (left) and large (right) cell with small dendrites, \textbf{(b)} the same with large dendrites. \textbf{(c-f)} Memory recall measured by the mutual information between the distribution of neuronal firing rates during learning and during recall stimulation (see \cite{LuboeinskiTetzlaff2021}), for a stimulated subset of varied size (a varied number of neurons are stimulated for learning/recall). Average over $100$ network realizations. Error bars represent the 95\% confidence interval. \textbf{(c)} Recall stimulation at $10\unit{s}$ after learning (technically, $D_\textrm{p}=10^{-11}\unit{m\textsuperscript{2}/s}$, but late-phase plasticity does not occur at such fast timescales). \textbf{(d)} Recall stimulation at $8\unit{h}$ after learning, $D_\textrm{p}=10^{-11}\unit{m\textsuperscript{2}/s}$. \textbf{(e)} Recall stimulation at $8\unit{h}$ after learning, $D_\textrm{p}=10^{-15}\unit{m\textsuperscript{2}/s}$. \textbf{(f)} Recall stimulation at $8\unit{h}$ after learning, $D_\textrm{p}=10^{-19}\unit{m\textsuperscript{2}/s}$. \textbf{(g-j)} Same as (c-f), but for large cells that consist of segments of twice the diameter.}
\\

\noindent {\small Figure S12: \textbf{Benchmarking results of total runtime per spike}, for 10s-recall paradigm in networks of $2000$ neurons, related to Arbor simulations in \red{Fig.~6} in the main article. \textbf{(a)} Total runtime (including setup and propagation phase) per spike, accounting for the fact that the spike numbers in the single-compartment and multi-compartment simulations are different (cf. panel (b)). The data points show the runtime as given in \red{Fig.~6a} in the main article, divided by the total number of spikes. \textbf{(b)} Total number of spikes in the different simulation paradigms. Data points represent the average over $10$ trials; error bars represent the standard deviation.}
\\

\noindent {\small Figure S13: \textbf{Benchmarking results of total runtime and memory use for 8h-recall paradigm.} Networks of $2000$ neurons (also see \red{Fig.~6} in the main article) in Arbor. The single-compartment simulations (`SC') are conducted as described in \red{subection 3.5} in the main article. The simulations with multi-compartment/morphological neurons (`MC') are conducted as described in \red{subsection 3.6}. Results are given for execution with and without SIMD support on an older desktop computer (Intel Core i5-6600 CPU @ 3.30GHz, $1\times8\unit{GB}$ DDR3-RAM, using $1$ thread). \textbf{(a)} Total runtime of the simulations, including setup and propagation phase. Measurements were performed using \texttt{hyperfine} in version 1.15. \textbf{(b)} Memory consumption, given by the maximum over time of the number of `dirty' bytes, including private and shared memory, as returned from the \texttt{pmap} tool. Data points represent the average over $10$ trials; error bars represent the standard deviation.}
\\

\noindent {\small Figure S14: \textbf{Benchmarking of total memory use for large-scale networks.} Total memory use to initialize and execute a simulation with $32768$ cells over $200\unit{ms}$ in Arbor and CoreNEURON. A busyring network of \texttt{simple-branchy} cells with tree depth $2$ is used, run on the HWS2 system (AMD Ryzen Threadripper PRO 5995WX CPU with 64 cores, $8\times32\unit{GB}$ DDR4-RAM) with \textbf{(a)} CoreNEURON, \textbf{(b)} Arbor with SIMD, \textbf{(c)} Arbor with SIMD with additional STDP mechanisms for the random synapses. The respectively lowest memory use for each implementation is highlighted by the gray box. \textbf{(d)} Scaling of the best results over network size. \textbf{(e)} Sketch of the busyring network consisting of rings of integrate-and-fire neurons (shown as blue disks), connected internally via excitatory synapses, and across the whole network via random synapses of weight zero. One neuron of each ring receives external stimulation. All values are averaged over $10$ trials, with coefficient of variation $\mathrm{CV} < 0.004$ in all cases. See Table~S1 for more details.}
\\

\noindent {\small Figure S15: \textbf{Strong scaling on the JUWELS and JEDI supercomputing systems.} %
Benchmarking results for the total wallclock time, including setup and state propagation phases, of a simulation with $10^6$ \texttt{complex} cells over $200\unit{ms}$.
We show strong scaling, i.e., the cell count is subdivided across the available GPUs. For reference, the intervals of 80\% efficiency are shown, where 100\% efficiency represents perfect scaling with the additional hardware. 
The data was obtained during the pre-production phase of the JUPITER supercomputing system on the JEDI preview system. Parameter values are provided in Table \red{7} in the main article; further parameters and hardware characteristics are provided in \red{Table~S3}.}
\\

\noindent {\small Table S1: \textbf{Total memory use for busyring benchmark.} 
Measurements are provided as reported by Arbor and CoreNEURON. Results are collected with the HWS2 system (AMD Ryzen Threadripper PRO 5995WX CPU with 64 cores, $8\times32\unit{GB}$ DDR4-RAM, without GPU). All values are averaged over $10$ trials, with coefficient of variation $\mathrm{CV} < 0.004$ in all cases. Arbor with SIMD. In CoreNEURON, the most extensive simulation did not finish (d.n.f.) due to exceeded memory. For the corresponding runtime results, cf. Table~\red{8} in the main article.}
\\

\noindent {\small Table S2: \textbf{Wallclock time measurements for busyring benchmark with different cell types.} Total runtime results are provided as reported by Arbor. All networks consist of 1024 neurons and 1.03 million synapses. The shares of the setup and state propagation phases are given in brackets, respectively. Results are collected with the HWS2 system (AMD Ryzen Threadripper PRO 5995WX CPU with 64 cores, $8\times32\unit{GB}$ DDR4-RAM, without GPU). All values are averaged over $10$ trials, with coefficient of variation $\mathrm{CV} < 0.06$ in all cases. Arbor with SIMD. Also see Tables~\red{8} and \red{9} in the main article for other paradigms.}
\\

\noindent {\small Table S3: \textbf{Parameters and hardware characteristics for investigating strong scaling on supercomputing systems.} 
We use one MPI rank per GPU and the full set of $4$ GPUs per node when present in the node architecture. See Table~\red{7} in the main article for further parameter values.}

\end{document}